\documentclass[prd, twocolumn, showpacs, superscriptaddress, floatfix, nofootinbib]{revtex4-1}

\usepackage[dvips]{graphicx}
\usepackage{epsf}
\usepackage{amsmath}
\usepackage{amssymb}

\usepackage{graphicx}
\usepackage{dcolumn}
\usepackage{bm}
\def\be{\begin{equation}}
\def\ee{\end{equation}}
\def\bea{\begin{eqnarray}}
\def\eea{\end{eqnarray}}

\usepackage{color}

\begin{document}


\title{Towards a Nonsingular Bouncing Cosmology}

\author{Yi-Fu Cai}
\email{ycai21@asu.edu}
\affiliation{Department of Physics, Arizona State University, Tempe, AZ 85287, USA}
\author{Damien A. Easson}
\email{easson@asu.edu}
\affiliation{Department of Physics, Arizona State University, Tempe, AZ 85287, USA}
\author{Robert Brandenberger}
\email{rhb@physics.mcgill.ca}
\affiliation{Department of Physics, McGill University, Montr\'eal, QC, H3A 2T8, Canada}


\begin{abstract}

We present a nonsingular bouncing cosmology using single scalar field matter with non-trivial potential and non-standard kinetic term. The potential sources a dynamical attractor solution with Ekpyrotic contraction which washes out small amplitude anisotropies. At high energy densities the field evolves into a ghost condensate, leading to a nonsingular bounce. Following the bounce there is a smooth transition to standard expanding radiation and matter dominated phases. Using linear cosmological perturbation theory we track each Fourier mode of the curvature fluctuation throughout the entire cosmic evolution. Using standard matching conditions for nonsingular bouncing cosmologies we verify that the spectral index does not change during the bounce. We show there is a controlled period of exponential growth of the fluctuation amplitude for the perturbations (but not for gravitational waves) around the bounce point which does not invalidate the perturbative treatment. This growth induces a natural suppression mechanism for the tensor to scalar ratio of fluctuations. Moreover, we study the generation of the primordial power spectrum of curvature fluctuations for various types of initial conditions. For the pure vacuum initial condition, on scales which exit the Hubble radius in the phase of Ekpyrotic contraction, the spectrum is deeply blue. For thermal particle initial condition, one possibility for generating a scale-invariant spectrum makes use of a special value of the background equation of state during the contracting Ekpyrotic phase. If the Ekpyrotic phase is preceded by a period of matter-dominated contraction, the primordial power spectrum is nearly scale-invariant on large scales (scales which exit the Hubble radius in the matter-dominated phase) but acquires a large blue tilt on small scales. Thus, our model provides a realization of the ``matter bounce" scenario which is free of the anisotropy problem.

\end{abstract}

\maketitle

\newcommand{\eq}[2]{\begin{equation}\label{#1}{#2}\end{equation}}

\section{Introduction}

In this paper we construct a nonsingular bouncing cosmology with an Ekpyrotic contracting phase. The model is free from the problems typically associated with `matter bounce' or pure Ekpyrotic models. Both the matter bounce (see e.g. \cite{RHB2011rev} for a recent review) and the Ekpyrotic scenario (see e.g \cite{Lehners} for an in-depth overview) were proposed as alternatives to inflationary cosmology as an explanation for the origin of the observed structure in the Universe. While these models are faced with serious challenges we show that these challenges can be ameliorated by combining attributes of both scenarios.

We begin with a discussion of the successes and problems of the matter bounce model. Some time ago it was realized that curvature fluctuations which originate as quantum vacuum fluctuations on sub-Hubble scales and which exit the Hubble radius during a matter-dominated epoch of contraction acquire a scale-invariant spectrum \cite{Wands, Fabio1}. In a number of toy models, it was shown that this spectrum often persists if the matter-dominated phase of contraction is continued in a nonsingular way to the expanding phase of Standard Big Bang cosmology. A nonsingular bouncing model can be obtained by using Null Energy Condition (NEC) violating matter such as quintom matter \cite{Yifu1, Yifu2} or Lee-Wick matter \cite{Yifu3}, by making use of either a ghost condensate construction \cite{Chunshan}, the ghost-free higher derivative gravity model of \cite{Tirtho1, Tirtho3}, or Galileon fields \cite{Taotao, Damien}, and it arises in Horava-Lifshitz gravity \cite{Horava} in the presence of spatial curvature \cite{HLbounce, HLbounce2}. A nonsingular bounce can also be obtained \cite{Omid, Easson:2007fz, Sahni:2012er} in the context of mirage cosmology \cite{Kehagias:1999vr, Shtanov:2002mb}, by taking into account the effect of extra time-like dimension \cite{Battefeld:2005cj}, by making use of nonconventional K-essence model \cite{Abramo:2007mp, Chimento:2005ua}, or in a universe with nontrivial curved geometries \cite{Martin:2003sf, Solomons:2001ef, Anabalon:2012tu}. In all of these cases it was found (under certain assumptions) that the spectrum of perturbations on scales relevant to current cosmological observations does not change its spectral index during the bouncing phase. Thus, such nonsingular cosmologies provide an alternative to inflation for producing a scale-invariant spectrum of cosmological
perturbations today.

A difficult problem facing bouncing cosmologies is that the homogeneous and isotropic background cosmological solution is unstable to the development of radiation \cite{Johanna} and anisotropic stress. The latter instability leads to the famous BKL \cite{BKL} mixmaster cosmology, opposed to a homogeneous and
isotropic bounce.

In the Ekpyrotic scenario \cite{Ekp}, the contracting branch solution is a local attractor \cite{Ekpattr}, similar to the accelerating expanding cosmological solutions in inflationary models. In Ekpyrotic cosmology there is the assumed existence of a matter fluid with an equation of state $w = p / \rho \gg 1$ ($p$ and $\rho$ being pressure and energy densities, respectively). The energy density in this fluid blueshifts faster than the contribution of anisotropic stress in the effective energy density. Thus, the fluid comes to dominate during the contracting phase and prevents the development of the BKL instability.

Within the context of a pure Ekpyrotic model, the curvature fluctuations on
super-Hubble scales in the contracting phase are not scale-invariant; although,
the fluctuation in the Bardeen potential $\Phi$ \cite{Bardeen}
\footnote{See \cite{MFB} for an in-depth treatment of the theory of
cosmological perturbations and \cite{RHBrev} for an introductory
overview.} which describes the metric
inhomogeneities in the longitudinal gauge are scale-invariant (see e.g.
\cite{Lyth, Fabio2, Hwang, KOST2} for discussions of this issue).
The initial Ekpyrotic scenario \cite{Ekp} had
a singularity in the effective field theory at the transition point between
contraction and expansion. The question of how fluctuations transfer from
the contracting to the expanding phase is non-trivial in this
context (see e.g. \cite{Durrer} for a detailed discussion). There are
prescriptions according to which the spectrum of fluctuations after
the bounce is scale-invariant \cite{Tolley}. Also, there are typically
entropy modes present in Ekpyrotic models. These entropy modes
can acquire a scale-invariant spectrum by the same mechanism that
$\Phi$ acquires such a spectrum
\cite{Notari, Finelli, Creminelli, Turok, Khoury}, and will then induce
a scale-invariant spectrum for the curvature fluctuations by the usual
mechanism of isocurvature modes seeding an adiabatic fluctuation. This
leads to the so-called ``New Ekpyrotic" scenario \cite{Khoury}.
Scale-invariant entropy modes arise  from fluctuations
of the extra metric fluctuation modes in higher dimensions \cite{Thorsten},
the setting in which the Ekpyrotic scenario was initially proposed.
However, in the context of the New Ekpyrotic scenario, one has to
worry about instabilities during the bounce phase,
an issue recently raised in \cite{BingXue}.

The model we propose in this paper is free from the complications
associated with entropy modes and the transfer of fluctuations through
singular bounces. In this scenario, it is the curvature fluctuations
which are scale-invariant. This spectrum is inherited from the phase
of matter contraction which preceded the phase of Ekpyrotic contraction.
There are no low mass entropy modes leading to
the complicated issues existing in pure Ekpyrotic models. The
cosmological bounce is nonsingular and hence the transfer of
fluctuations from the initial contracting phase to the expanding period
can be treated exactly. We find that the spectral index of the
fluctuations on infrared scales relevant to current observations
does not change during the bounce. The amplitude of the curvature
fluctuations are boosted by a significant, scale-independent, factor ${\cal F}$,
(for long wavelength
modes). We demonstrate that our model is free from
the instability problems raised in \cite{BingXue}.

In this model the background dynamics before and during the bounce
are determined by a single scalar field $\phi$ with non-trivial kinetic action of
the form used in ghost condensate \cite{ghost} and Galileon models
\cite{Galileon} (see also \cite{Fairlie}). We add to this
field a negative exponential potential similar to what is used in the Ekpyrotic
scenario. This potential leads to a phase of Ekpyrotic contraction.
At high densities, the coefficient of the part of the kinetic term which is standard becomes negative, and this leads to a nonsingular cosmological
bounce similar to what is achieved in the New Ekpyrotic scenario
\cite{Khoury, Creminelli} and in \cite{Chunshan}. After the bounce, the
coefficient of the kinetic term becomes positive again, a
period of kinetic-driven expansion sets in, lasting until the radiation and
matter dominated periods of Standard Cosmology.

We can imagine that the initial contracting phase of the universe
mimics our currently observed universe,
namely a state filled with regular matter and radiation. Since the kinetic
energy density of $\phi$ increases faster in the contracting phase than
both that of matter and radiation, eventually
$\phi$ will begin to dominate. Given this setup, our model provides a realization
of the matter bounce scenario of \cite{Wands, Fabio1, RHB2011rev}
which is free from the anisotropy problem which generically affects
bouncing cosmologies.

The outline of this paper is as follows: In the next section we introduce
the model and discuss the cosmological background dynamics, both
analytically and numerically. In Section III we show how
cosmological fluctuations evolve through the bouncing phase. The final section
is reserved for discussion.

We defined the reduce Planck mass by $M_{p} = 1/\sqrt{8\pi G}$ where $G$
is Newton's gravitational constant. The sign of the metric is taken to be
$(+,-,-,-)$. Note that we take the value of the scale factor at the
bounce point to be $a_B = 1$ throughout the paper.

\section{Model and Background Cosmology}

It is well known that, in order to realize a successful nonsingular
homogeneous and isotropic bounce in a
spatially flat ($k=0$) Friedmann-Robertson-Walker-Lemaitre (FRW) universe in the framework
of standard Einstein gravity, the Null Energy Condition for matter must be violated.
This is because at the bounce point the time derivative of the Hubble parameter is greater than zero while the Hubble parameter itself is zero. This implies that the
total energy density vanishes while the pressure is negative, \it i.e.\rm, that
the background equation of state is $w < -1$. One way to achieving such a
scenario is via a ghost condensate field $\phi$ \cite{Khoury, Creminelli, Chunshan}
in which the Lorentz symmetry is broken spontaneously in the infrared, with  the kinetic term
for $\phi$ taking on a non-vanishing expectation value. However, this type of model
suffers a gradient instability since the square of the sound speed of its perturbations
becomes negative in the phase of ghost condensation \footnote{If the period of ghost
condensation is short, as it is in the model of \cite{Chunshan}, the instabilities do not
have sufficient time to grow to a damaging magnitude.}.  Another mechanism of achieving
Null Energy Condition violation is to make use of a Galileon type field. The key feature
of this type of field is that it contains higher order derivative terms in the Lagrangian
while the equation of motion remains second order, and thus does not lead to the appearance of
ghost modes. Galileon fields have been used to construct emergent universe scenarios
in which the universe begins with a quasi-static phase \cite{Nicolis}. There have also
been recent studies of Galileon models which yield cosmologies with a nonsingular
bounce, but in which a space-time singularity of big rip type arises after the bouncing phase
\cite{Taotao, Damien}. In this section, we present a nonsingular bouncing solution in
terms of a single field having the desirable features of both the ghost condensate
and Galileon-inspired models.

\subsection{The Model}

The most general form of single scalar field Lagrangian giving rise to second-order
field equations in  four-dimensional spacetime can be expressed
as~\cite{Horndeski:1974, Deffayet:2011gz}
\begin{eqnarray}\label{Lagrangian}
 {\cal L} \, = \, K(\phi, X) + G(\phi, X) \Box \phi + L_4 + L_5 ~,
\end{eqnarray}
where $K$ and $G$ are functions of a dimensionless scalar field $\phi$ and its
canonical kinetic term
\begin{equation}
X \equiv \partial_\mu \phi \partial^\mu \phi/2 \, .
\end{equation}
The standard kinetic Lagrangian corresponds to $K = X$ and all other terms
vanishing. A more general form is $K(\phi, X) = A(\phi) B(X)$
with $A(\phi) > 0$ used in  ``K-essence" models \cite{kessence}.

The other kinetic terms of $\phi$ include the operator
\begin{equation}
\Box\phi \equiv g^{\mu\nu} \nabla_\mu \nabla_\nu \phi \, .
\end{equation}
The terms $L_4$ and $L_5$ are higher order operators which are usually suppressed
at low energy scales. Thus, in the present paper we will ignore them and focus on the
first two terms adopting the form in the ``Kinetic Gravity Braiding" (KGB)
model of \cite{Deffayet:2010qz} (see also \cite{Yokoyama}).

Variation of the above matter action minimally coupled to Einstein gravity leads to the
modified Einstein equation:
\begin{eqnarray}\label{energystress}
 T_{\mu\nu} &=& M_p^2 \left(R_{\mu\nu}-\frac{R}{2}g_{\mu\nu} \right) \nonumber\\
 &=& (-K+2XG_{,\phi}+G_{,X}\nabla_\sigma{X}\nabla^\sigma\phi)g_{\mu\nu} \nonumber\\
 && + (K_{,X}+G_{,X}\Box\phi-2G_{,\phi})\nabla_\mu\phi\nabla_\nu\phi \nonumber\\
 && - G_{,X}(\nabla_\mu{X}\nabla_\nu\phi+\nabla_\nu{X}\nabla_\mu\phi)~.
\end{eqnarray}
In the above formalism, $_{,\phi}$ and $_{,X}$ denote derivatives with respect to
$\phi$ and $X$, respectively.

For the model under consideration we choose:
\begin{eqnarray}\label{Kessence}
 K(\phi, X) = M_p^2 \left[1-g(\phi) \right]X + \beta X^2 - V(\phi)~,
\end{eqnarray}
where we introduce a positive-definite parameter $\beta$ so that the kinetic term is bounded
from below at high energy scales. For $g > 1$ a ghost condensate ground state with
$X \neq 0$ can arise. Note that the first term of $K$ involves $M_p^2$ since in
the present paper we adopt the convention that the scalar field $\phi$ is dimensionless.
We have also introduced a non-trivial potential $V$ for $\phi$. This potential is chosen
such that Ekpyrotic contraction is possible. In the specific model which we will discuss
in the following, the scalar field evolves monotonically from a negative large value to a
positive large value. The function $g(\phi)$ is chosen such that a phase of ghost
condensation only occurs during a short time when $\phi$ approaches $\phi=0$. This
requires the dimensionless function $g$ to be smaller than unity when $|\phi| \gg 1$
but larger than unity when $\phi$ approaches the origin.

The term $G$ is a Galileon type operator which  is consistent with the fact that
the Lagrangian contains higher order derivative terms in $\phi$, but the equation
of motion remains a second order differential equation. Phenomenologically, there are
few requirements on the explicit form of $G$. We introduce this operator since we expect
that it can be used to stabilize the gradient term of cosmological perturbations, which
requires that the sound speed parameter behaves smoothly and is positive-definite
throughout most of the background evolution. For simplicity, we will choose $G$
to be a simple function of only $X$:
\begin{eqnarray}\label{Galileon}
 G(X) = \gamma X~,
\end{eqnarray}
where $\gamma$ is a positive-definite number.

\subsection{The Cosmological Background Evolution}

We now study the equations of motion which result from
applying the Lagrangian (\ref{Lagrangian}) to the flat FRW universe whose
metric is given by
\begin{eqnarray}\label{FRWmetric}
 ds^2 \, = \, dt^2-a^2(t)d\vec{x}^2~,
\end{eqnarray}
where $t$ is cosmic time, $x$ are the comoving spatial coordinates and
$a(t)$ is the scale factor. Thus, the kinetic terms of a homogeneous and
isotropic scalar field background become
\begin{eqnarray}
 X &=& \dot\phi^2/2 ~, \nonumber \\
 \Box\phi &=& \ddot\phi + 3H\dot\phi ~.
\end{eqnarray}
The dot denotes the derivative with respect to  cosmic time $t$. The Hubble
parameter is $H \equiv \dot{a}/a$.

{F}or this background the energy density is
\begin{eqnarray}\label{rho}
 \rho = \frac{1}{2}M_p^2 (1-g)\dot\phi^2 +\frac{3}{4}\beta\dot\phi^4 +3\gamma H\dot\phi^3 +V(\phi)~,
\end{eqnarray}
and the pressure is
\begin{eqnarray}\label{pressure}
 p = \frac{1}{2}M_p^2 (1-g)\dot\phi^2 +\frac{1}{4}\beta\dot\phi^4 -\gamma\dot\phi^2\ddot\phi -V(\phi)~,
\end{eqnarray}
by computing the diagonal components of the stress-energy tensor (\ref{energystress}).

To derive the equation of motion for $\phi$, one can either vary the Lagrangian with respect
to $\phi$ or require that the covariant derivative of the stress-energy tensor vanish.
This yields
\begin{eqnarray}\label{eom}
 {\cal P} \ddot\phi + {\cal D} \dot\phi +V_{,\phi} = 0~,
\end{eqnarray}
where we have introduced
\begin{eqnarray}
\label{Pterm}
 {\cal P} &=& (1-g)M_p^2 +6\gamma H\dot\phi +3\beta\dot\phi^2 +\frac{3\gamma^2}{2M_p^2}\dot\phi^4~,\\
\label{Fterm}
 {\cal D} &=& 3(1-g)M_p^2H +(9\gamma{H}^2-\frac{1}{2}M_p^2g_{,\phi})\dot\phi +3\beta{H}\dot\phi^2 \nonumber\\
 &&-\frac{3}{2}(1-g)\gamma\dot\phi^3 -\frac{9\gamma^2H\dot\phi^4}{2M_p^2} -\frac{3\beta\gamma\dot\phi^5}{2M_p^2}~.
\end{eqnarray}
The coefficient ${\cal P}$ determines the positivity of the kinetic term of the scalar field
and thus can be used to judge whether the model contains a ghost or not at the perturbative
level. The coefficient ${\cal D}$ is an effective damping term. By keeping the first terms
of the expressions of ${\cal P}$ and ${\cal D}$ and setting $g = 0$ one can recover the
standard Klein-Gordon equation in the FRW background. Neglecting the other terms is a good approximation when the velocity of $\phi$ is sub-Planckian.

\subsection{The Nonsingular Bounce Solution}

We now consider a nonsingular cosmological bouncing solution.
Here, we focus on homogeneous solutions.
It is well known that the homogeneous trajectory of a scalar field can be an attractor
solution when its potential is an exponential function. One example is inflationary
expansion of the universe in a positive-valued exponential potential, and the
other one is the Ekpyrotic model in which the homogeneous field trajectory for a
negative exponential potential is an attractor in a contracting universe. For
a phase of Ekpyrotic contraction, we take the form of the potential to be
\begin{eqnarray}
 V(\phi) = -\frac{2V_0}{e^{-\sqrt{\frac{2}{q}}\phi}+e^{b_V\sqrt{\frac{2}{q}}}\phi}~,
\end{eqnarray}
where $V_0$ is a positive constant with dimension of $({\rm mass})^4$. Thus the
potential is always negative and asymptotically approaches zero when
$|\phi| \gg1$. Ignoring the second term of the denominator, this potential reduces to
the form used in the Ekpyrotic scenario.

For initial conditions we assume that $\phi$ begins at an asymptotically large
negative value. The force due to the potential  induces motion towards
the right, i.e. with ${\dot \phi} > 0$. The equation of state of $\phi$ matter is
determined by the parameter $q$. For sufficiently small values of $q$ (as
will be shown shortly, the transition value is $q = 1/3$), the equation of
state is such that in a contracting universe the energy density in $\phi$
increases faster than that in matter, radiation and anisotropic stress.
Thus, the homogeneous trajectory is an attractor. As $\phi$ approaches
zero, the second  term in the denominator of the potential becomes important and
the field evolution departs from the Ekpyrotic trajectory. Since the potential is
bounded from below the model has a stable vacuum state.

To obtain a nonsingular bounce, we must make an explicit choice of
$g$ as a function of $\phi$. As we have discussed in the previous subsection, we want
$g$ to be negligible when $|\phi|\gg1$. In order to obtain a violation of the Null
Energy Condition after the termination of the Ekpyrotic contracting phase, $g$ must
become the dominant coefficient in the quadratic kinetic term when $\phi$ approaches $0$.
Thus, we suggest its form to be
\begin{eqnarray}
 g(\phi) = \frac{2g_0}{e^{-\sqrt{\frac{2}{p}}\phi}+e^{b_g\sqrt{\frac{2}{p}}\phi}}~,
\end{eqnarray}
where $g_0$ is a positive constant defined as the value of $g$ at the moment of
$\phi=0$, which is required to be larger than $1$ as will be discussed later.

\begin{figure}
\includegraphics[scale=0.3]{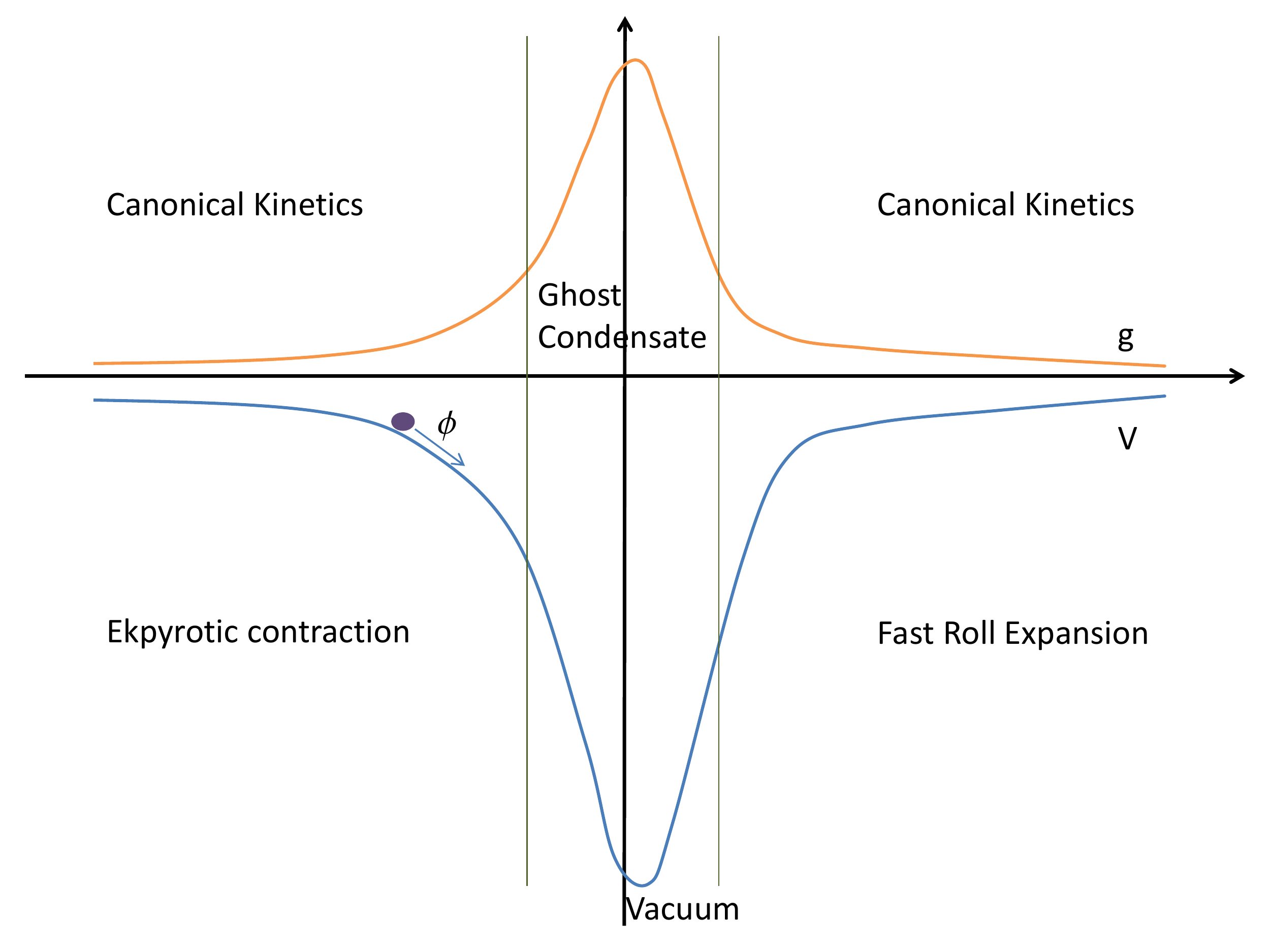}
\caption{A sketch of the coefficient $g$ and of the potential $V$ as functions
of the scalar field $\phi$. Depending on the value of $g$, the field space is
separated into three regimes: the contracting phase with nearly canonical kinetic
term (large negative field values), the ghost condensate phase when $g > 1$ which
occurs at field values close to $0$, and the expanding phase with canonical kinetic
term (large positive field values), respectively. We
find that the phase of ghost condensation leads to a nonsingular
cosmological bounce.}
\label{Fig:sketch}
\end{figure}

At this point we have fully determined the model. The potential $V(\phi)$ and
the function $g(\phi)$ are sketched as a function of $\phi$ in
Figure \ref{Fig:sketch}. The horizontal axis is the field value $\phi$, the
vertical axis shows the values of the functions $V$ (blue curve, negative-definite)
and $g$ (red curve, positive-definite). This figure is helpful in gaining
a semi-analytic understanding of the evolution, the topic we turn to in the
following subsection.

\subsection{Semi-Analytic Analysis}\label{Sec:semi-analytic}

As initial conditions for $t \ll -1$ (in Planck units or units of the typical
mass scale in the Lagrangian, whichever yields a larger time),
we assume a nearly homogeneous universe filled
with pressureless matter, radiation, and a homogeneous $\phi$ field
condensate with $\phi \ll -1$ which is in a contracting phase.
From Fig. \ref{Fig:sketch}, we see that in this phase $g \rightarrow 0$
and (since the gradient of the potential also approaches zero) $\dot\phi \ll M_p$.
In this case the Lagrangian for $\phi$ approaches the conventional canonical form:
\begin{eqnarray}\label{LagEk}
 {\cal L} \rightarrow \frac{M_p^2}{2}\partial_\mu\phi\partial^\mu \phi -V(\phi)~,
\end{eqnarray}
This Lagrangian admits the Ekpyrotic attractor solution for $\phi$ in a
contracting universe:
\begin{eqnarray}\label{phiEk}
 \phi_c \simeq -\sqrt{\frac{q}{2}} \ln \left[\frac{2V_0t^2}{q(1-3q)M_p^2} \right]~,
\end{eqnarray}
which yields an effective equation of state
\begin{eqnarray}\label{eos_Ek}
 w_c \simeq -1+\frac{2}{3q}~,
\end{eqnarray}
where the subscript ``$_c$" denotes the contracting phase. Consequently, long
before the bounce, the energy density of the scalar field $\phi$ evolves as that of
a perfect fluid with a constant equation of state $w_c$. Provided $w_c$ is the
largest among the equations of state of all the matter components in the universe,
the contribution of $\phi$ to the total energy density will become dominant,
as is well-known in the Ekpyrotic model. If $q < 1/3$ we have $w_c > 1$ and
in this case the importance of $\phi$ increases also relative to that of
anisotropic stresses, demonstrating that the model is free from the BKL
instability which plagues most bouncing cosmologies.

As $\phi$ accelerates towards $\phi = 0$, the value of $g$ will increase. If
$g_0 > 1$ (which we require), then at some point in time $g$ will start to
exceed the critical value $g = 1$ and thus the sign of quadratic kinetic term
in (\ref{Lagrangian}) will become negative. At that point, $\phi$  will become
a ghost condensate. The critical value of $g$ which signals the onset of
the ghost condensate phase is
\begin{eqnarray}
 g(\phi_{*}) = 1~.
\end{eqnarray}
There are two solutions of this equation which are
\begin{eqnarray}
\phi_{*-} &\simeq& -\frac{\ln{2g_0}}{p} \,\,\, {\rm and}  \nonumber \\
\phi_{*+} &\simeq& \frac{\ln{2g_0}}{b_gp} \, .
\end{eqnarray}
The ghost condensate phase occurs for $\phi_{*-} < \phi < \phi_{*+}$. During
this phase the Null Energy Condition is violated~\footnote{It is not sufficient to
have $\rho = 0$ which could occur for $\beta = 0$ before ghost condensation
sets in.}. This allows for the existence of a nonsingular bounce. However,
a nonsingular bounce also requires that the energy density vanishes at the bounce
point, which implies the following relation
\begin{eqnarray}
 \frac{1}{2}M_p^2 \left(1-g_B \right)\dot\phi_B^2 +\frac{3}{4}\beta\dot\phi_B^4 +V_B = 0~,
\end{eqnarray}
at the time $t_B$ when the bounce occurs (subscripts B refer to the bounce point).
In our specific example, $\phi_B \simeq 0$ and thus $g_B \simeq g_0$ and
$V_B \simeq V_0$. Therefore, we find that
\begin{eqnarray}\label{dotphiB}
 \dot\phi_B^2 \simeq \frac{(g_0-1)M_p^2}{3\beta} \left[ 1 +\sqrt{1 +\frac{12\beta{V}_0}{(g_0-1)^2M_p^4}} \,\right]~,
\end{eqnarray}
at the time of the bounce. From this result it follows immediately that a nonsingular
bounce can only occur when $g_0>1$ (otherwise there exists no real solution to
Eq. (\ref{dotphiB})).

Since it has a large positive velocity at the bounce point, $\phi$ continues
to increase after the bounce. Within a short time it will cross the second boundary
of the ghost condensation region $\phi_{*+}$. At that point, the Lagrangian of the
model recovers the canonical form and the universe enters a kinetic-driven
phase of expansion. Note that although the potential approaches an exponential form, the
scalar field does not approach the solution analogous to (\ref{phiEk}) which would
be an attractor for positive $\phi$ values in a contracting universe. In an
expanding space-time, this solution is a repeller rather than an attractor.
Instead, the scalar field experiences a fast rolling phase with an effective equation of
state the same as that of a stiff fluid:
\begin{eqnarray}\label{eos_FR}
 w_e \simeq 1~,
\end{eqnarray}
where the subscript ``$_e$" denotes the expanding phase. As a consequence,
the energy density of the scalar field $\phi$ will dilute relative to that of
conventional radiation and matter and then the universe will be able to enter
the phases of the usual thermal history of Standard Big Bang cosmology.
Following our qualitative analytical analysis we now turn to a numerical study of
the background cosmology.

\subsection{Numerical Analysis}

Above, we have presented a heuristic discussion of how the Null Energy Condition can
be violated in this setup. The key issue for the numerical analysis is to analyze
whether the Null Energy Condition can be violated smoothly and without any
pathologies. In our model, we make use of the idea of ghost condensation to violate
the Null Energy Condition. It is well known that any single field described
by a K-essence type Lagrangian \cite{kessence} will not cross the cosmological constant boundary (see e.g. Appendix of \cite{Xia}). For a  more general ghost condensate Lagrangian, violation of
the Null Energy Condition is possible, but perturbations could become very large
and force the background trajectory back away from the cosmological constant
boundary. In addition, there are gradient instabilities \cite{ghost}. To cure these, we take into account the effect of the Galileon operator $G(X)=\gamma X$.

In order to prove that the model is well-behaved throughout the entire cosmological
evolution, we need to study the evolution of perturbations about the background
solution. We perform the analysis of cosmological perturbation
in the next section.  Here we will focus on the pure background dynamics.

To illustrate that a nonsingular bounce can be achieved in our model,
we numerically evolve the Einstein acceleration equation coupled to the field equation for
$\phi$, imposing the Hamiltonian constraint equation to set the intial
conditions. In the numerical computations we
work in units of the Planck mass $M_p$ for all parameters. Specifically,
these parameters are chosen to be:
\begin{eqnarray}\label{parameters}
 &V_0 = 10^{-7}~,~~g_0 = 1.1~,~~\beta = 5~,~~\gamma = 10^{-3}~,\nonumber\\
 &b_V = 5~,~~b_g = 0.5~,~~p = 0.01~,~~q=0.1~.
\end{eqnarray}

\begin{figure}
\includegraphics[scale=0.3]{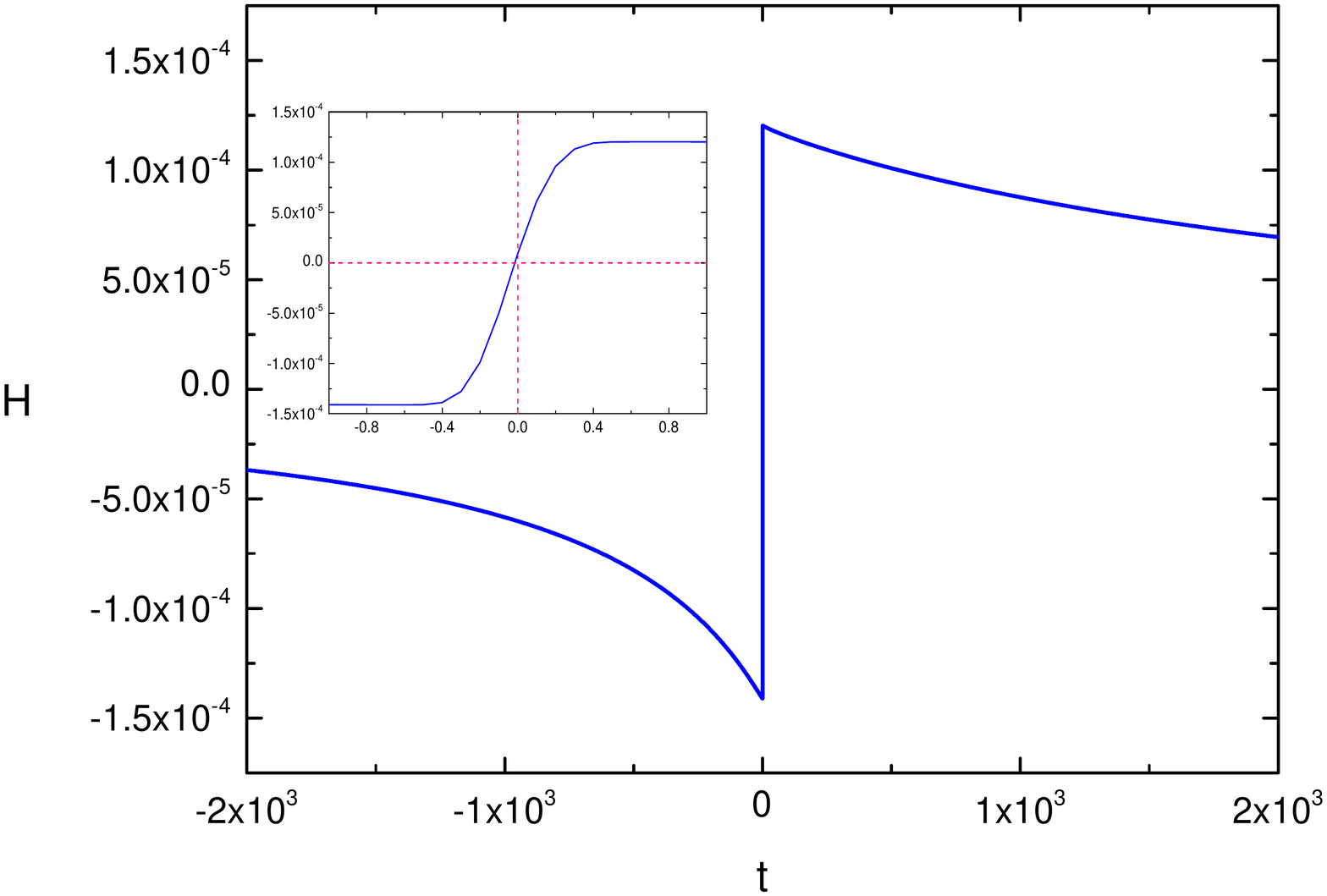}
\caption{Numerical plot of the Hubble parameter $H$ (vertical axis) as a function
of cosmic time (horizontal axis). The main plot shows that a nonsingular bounce occurs,
and that the time scale of the bounce is short (it is a ``fast bounce" model). The inner
insert shows a blowup of the smooth Hubble parameter $H$ during the bounce phase.
The background parameters were chosen as in Eq. (\ref{parameters}).
All numerical values are in Planck units $M_p$. The initial conditions were chosen
as described in the following figure. }
\label{Fig:Hubble}
\end{figure}

\begin{figure}
\includegraphics[scale=0.3]{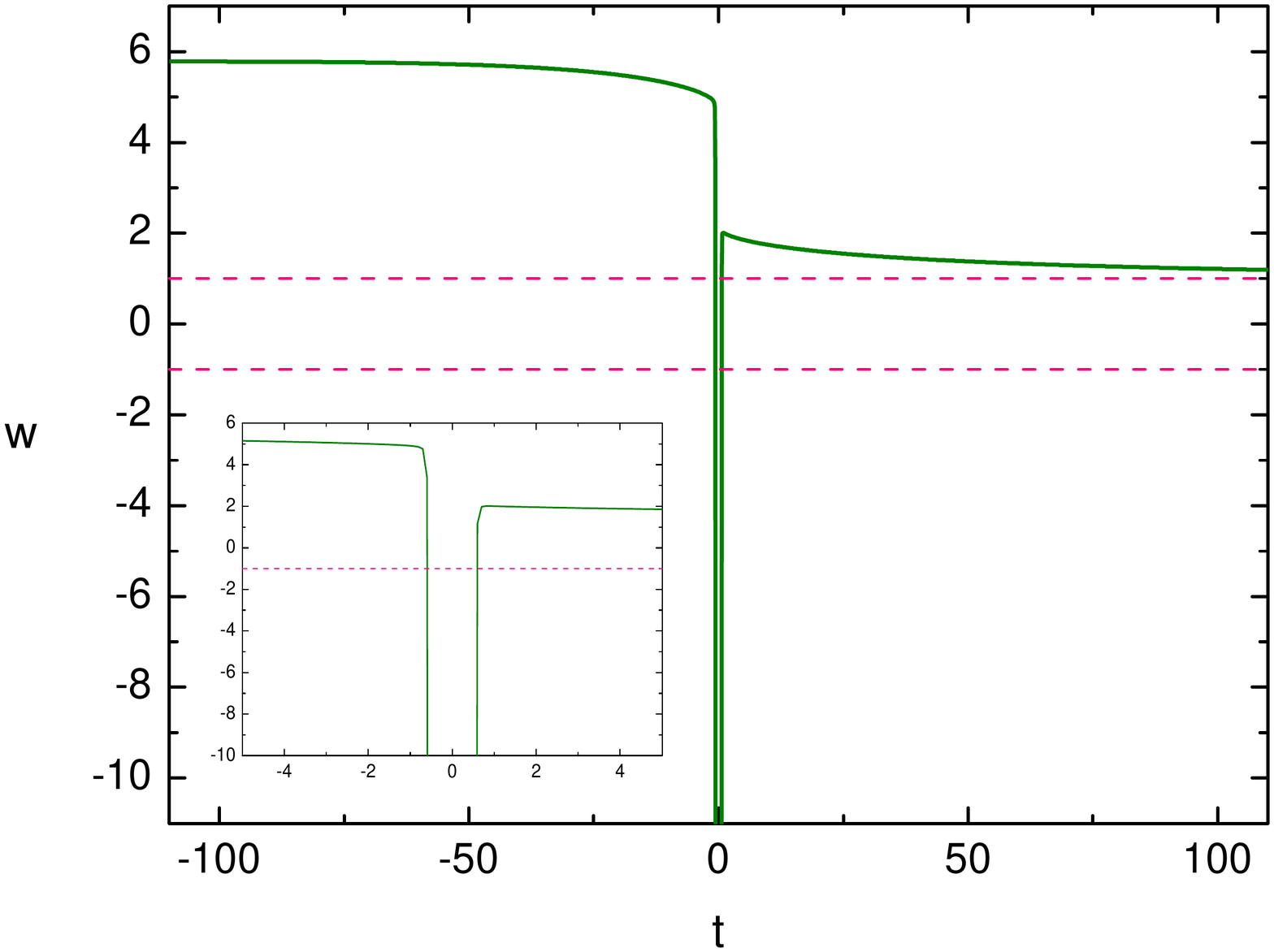}
\caption{Numerical plot of the evolution of the equation of state parameter $w$
(vertical axis) as  a function of cosmic time (horizontal axis). The insert shows
the detailed evolution of $w$ around the bounce time. The initial conditions were
chosen to be: $\phi_i = - 0.73$ and $\dot\phi_i = 1.5 \times 10^{-4}$. The background
parameters were chosen as listed in Eq. (\ref{parameters}). Planck units are used.}
\label{Fig:EoS}
\end{figure}

In Figs. \ref{Fig:Hubble}, and \ref{Fig:EoS}, respectively, we plot the
numerical results for the evolution of the Hubble parameter and the
equation of state. Also shown are zoomed-in views of the evolution around the bounce
point. One can see from Fig. \ref{Fig:Hubble} that the Hubble parameter $H$ evolves
smoothly through the bounce point with a dependence on cosmic time which
is close to linear \footnote{In the following section we make use of the linear
approximation for $H$ around the bounce point in our analytical study of the
evolution of cosmological fluctuations.}. The maximal value of the Hubble
parameter $H$, which we denote as the bounce scale $H_B$, is of the order
$O(10^{-4}M_p)$, and it is mainly determined by the value of the parameter $V_0$.

{F}rom Fig. \ref{Fig:EoS}, one sees that the equation of state parameter $w$ of the
scalar field is approximately equal to $w = 5.67$ which agrees well with what
is obtained by inserting $q=0.1$ into Eq. (\ref{eos_Ek}) \footnote{Since in the contracting
phase the solution of (\ref{eos_Ek}) is stable along with the background evolution,
one can choose any arbitrary value of $q$ to design a nonsingular bouncing model.}.
However, the Ekpyrotic contracting phase ends when $H$ approaches  the bounce
scale $H_B$. We see that $w$ then crosses the cosmological constant divide $w = -1$,
and this implies the violation of the Null Energy Condition in the bounce phase.
After the bounce, the equation of state rapidly evolves back to be above the
cosmological constant boundary and quickly approaches the
value $w = 1$ which corresponds to the kinetic-driven phase of expansion.
This fast-roll expanding phase is determined by the shape of the potential. During
this period  the contribution of the scalar field $\phi$ will be diluted quickly relative to
the contributions of regular matter and radiation. Thus, the universe in our model
is able to connect smoothly to the usual thermal history of the Standard Big
Bang model.

\begin{figure}
\includegraphics[scale=0.3]{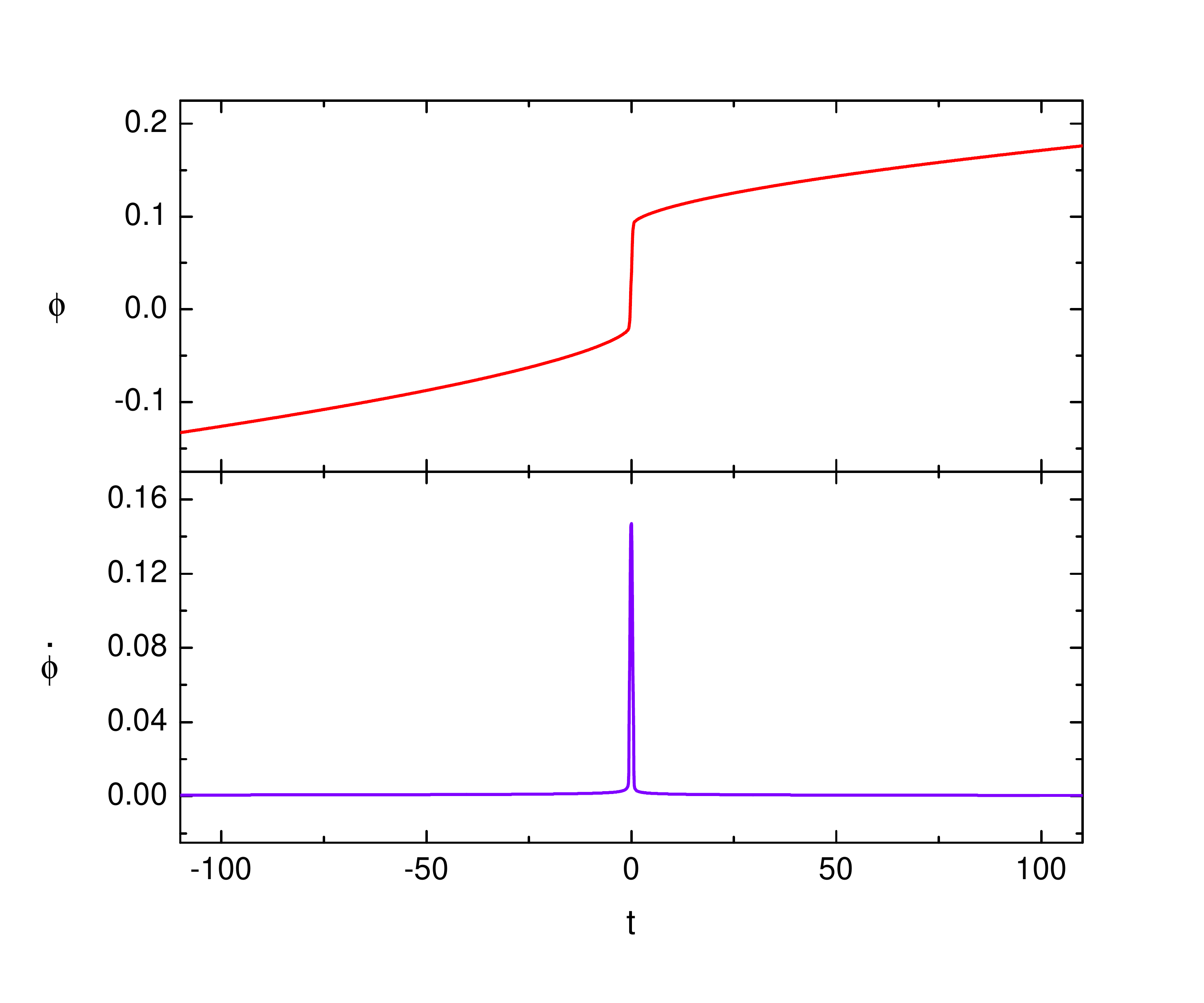}
\caption{Plot of the background scalar field $\phi$ and its time derivative $\dot\phi$
as functions of cosmic time. The initial conditions and background parameters are
the same as those chosen in Fig. \ref{Fig:Hubble}. Planck units are used. }
\label{Fig:phi}
\end{figure}

We also plot the numerical results of the evolution of the background scalar field
$\phi$ and its time derivative $\dot\phi$ as functions of cosmic time (see Fig. \ref{Fig:phi}).
We see that the scalar $\phi$ evolves monotonically from a large negative value
to a large positive one. Notice that far away from the bounce point, both the evolution
of $\phi$ and $\dot\phi$ are smooth, and  $\dot\phi \ll M_p$ which implies that the
higher order operators in the Lagrangian (\ref{Lagrangian}) are highly suppressed
by powers of $M_p$. This explains why the approximate Lagrangian (\ref{LagEk})
used in the semi-analytic analysis is valid. In the bounce phase,
$\dot\phi$ suddenly obtains a dramatic enhancement and then, after the bounce, falls
back to a small value. Correspondingly, there is a sharp peak in the plot of $\dot\phi$
as a function of time. This is related to a short-term tachyonic instability about the bounce
point which could lead to a ``big rip" singularity as was found in the cosmology
of the conformal Galileon \cite{Taotao, Damien}. However, since in the model the
potential is bounded from below and the bounce phase only lasts a couple of
Planck times, this instability does not have time to develop into something
which could destabilize the background solution.

\begin{figure}
\includegraphics[scale=0.3]{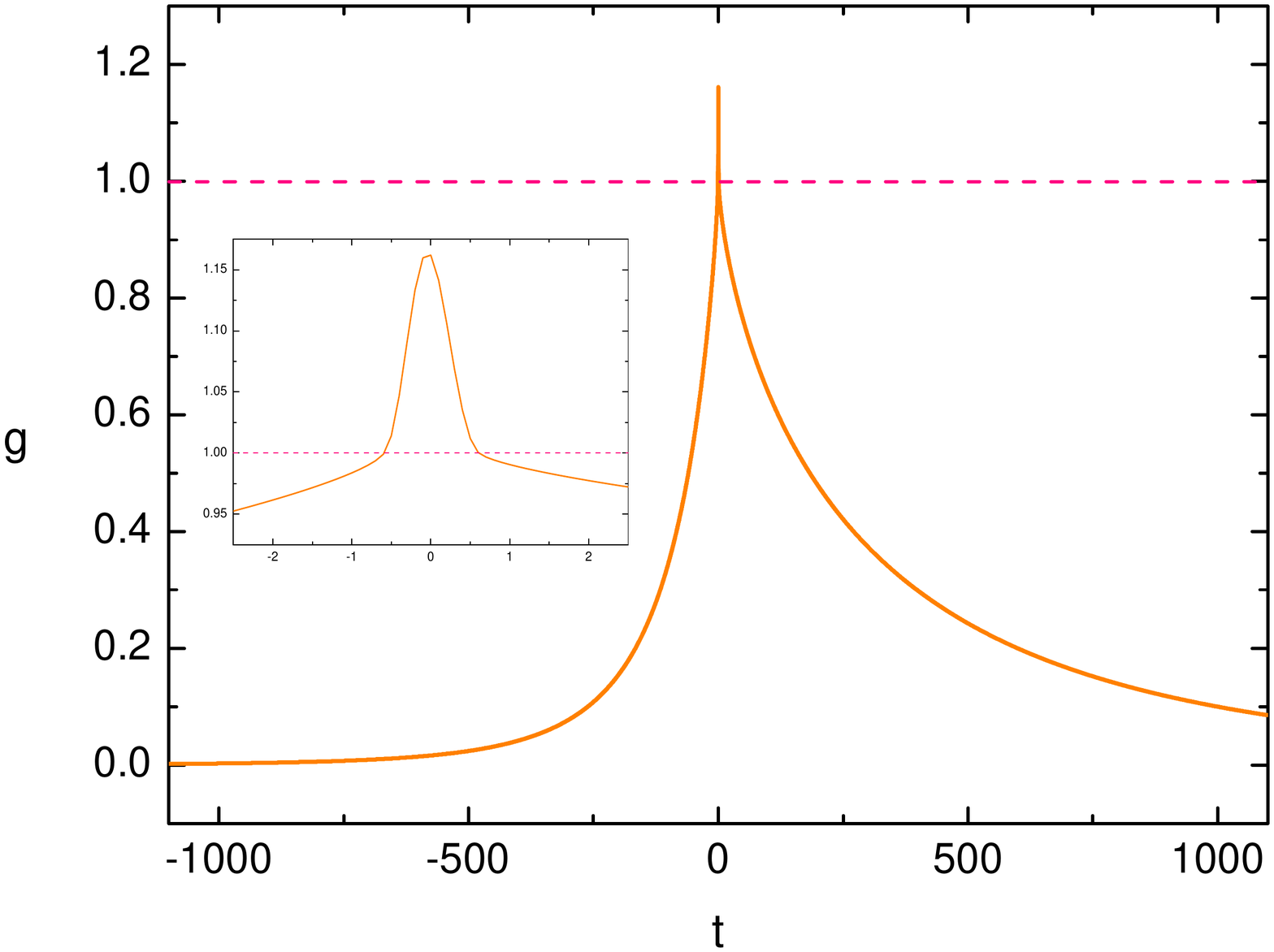}
\caption{Numerical plot of the ghost coefficient $g$ as a function of cosmic time. The insert
shows the detailed evolution of $g$ close to the bounce point. During the interval when
$g > 1$ a ghost condensate will have formed. The initial conditions and background
parameters are the same as those chosen in Fig. \ref{Fig:Hubble}. Again, Planck units
are used. }
\label{Fig:ghost}
\end{figure}

\begin{figure}
\includegraphics[scale=0.3]{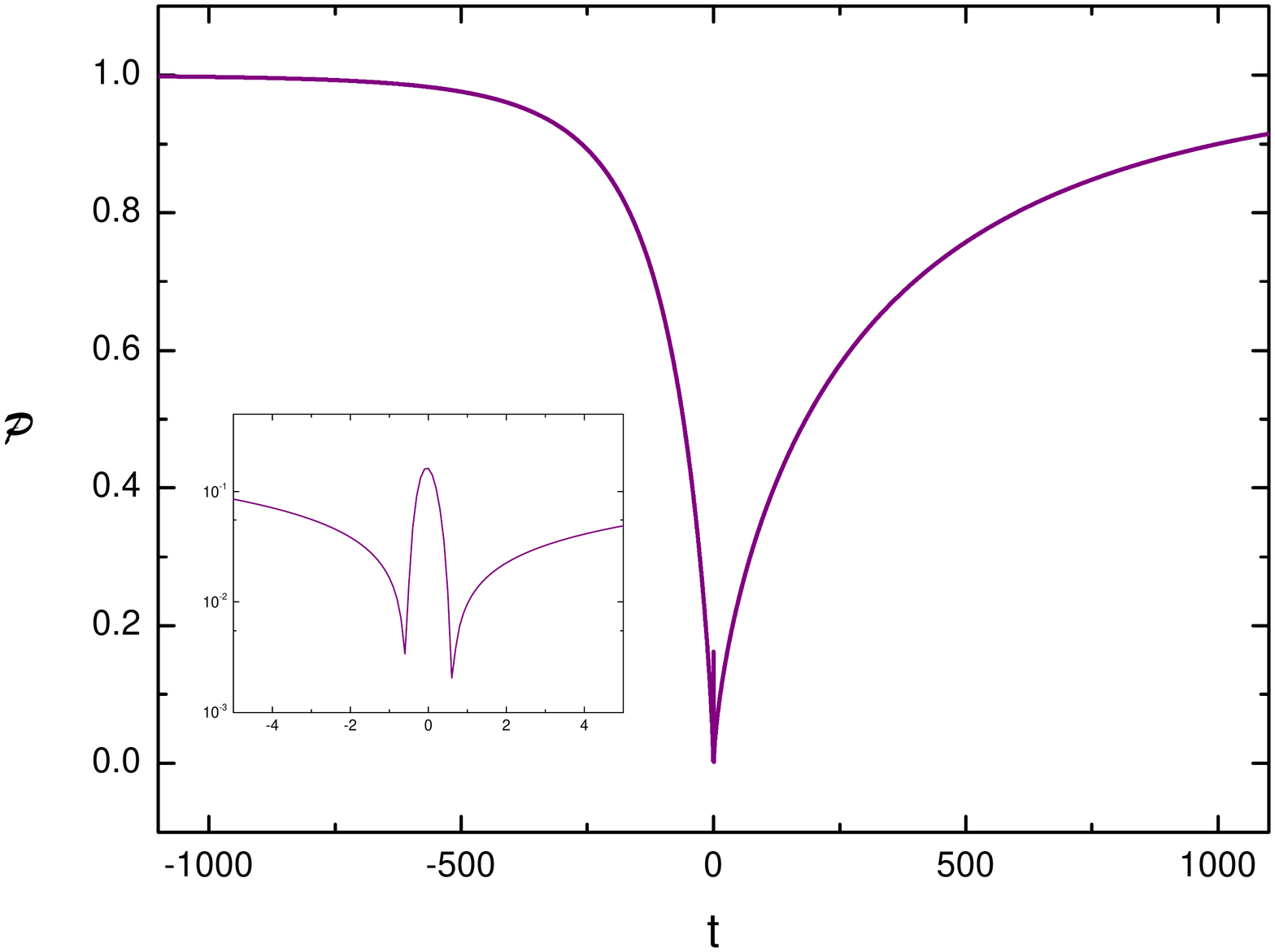}
\caption{Numerical plot of the coefficient ${\cal P}$ as a function of cosmic time. The insert
shows the detailed evolution of ${\cal P}$ near the bounce point. The initial conditions
and background parameters are the same as those chosen in Fig. \ref{Fig:Hubble}.
Planck units are used. Note that the coefficient is positive throughout which implies the
absence of ghost instabilities in the infrared limit. }
\label{Fig:Psign}
\end{figure}

{F}inally, we calculate the evolution of the coefficients $g$ and ${\cal P}$ (see
Figs. \ref{Fig:ghost} and \ref{Fig:Psign}). From Fig. \ref{Fig:ghost} We see that $g \ll 1$
far away from the bounce point, but becomes larger than one during the bounce phase
which indicates that there exists a period during which a ghost condensate forms. However,
the model is free of any ghosts in the far infrared since the sign of the coefficient
${\cal P}$ is always positive as shown in Fig. \ref{Fig:Psign}.


To conclude this subsection, we have verified that our model leads to a nonsingular
bounce which is free of instabilities at the level of homogeneous and isotropic
cosmology. The stability of the model towards inhomogeneities is investigated in the
next section.

\subsection{On the Stability of the Bounce}

In recent work \cite{BingXue}, Steinhardt and Xue raised important concerns regarding
the stability of modified Ekpyrotic bounces. They find that in the model \cite{Khoury}
the anisotropy and the shear which are damped out
during the phase of  Ekpyrotic contraction shoot up to values larger than their initial values
during the nonsingular bounce phase. In addition, they find that the adiabatic mode which is
suppressed during the phase of Ekpyrotic contraction but which has a deep blue
spectrum increases dramatically in amplitude during the bounce phase and
dominates over the entropy mode which in the construction of \cite{Khoury} has
a scale-invariant spectrum.

As we discuss in this subsection, our model appears to be free of these problems
\footnote{We are grateful to Paul Steinhardt and BingKan Xue for detailed
discussions on this point.}. The model of \cite{Khoury} also uses a ghost
condensate construction to achieve a nonsingular bounce. However, the
Null Energy Condition-violating ghost condensate appears at small values of $X$, namely
for $X < X_c$, where $X_c$ is some critical value. In the model of \cite{Khoury},
the value $X_i$ of $X$ at the beginning of the Ekpyrotic phase of contraction is
larger than $X_c$, and it increases rapidly during the Ekpyrotic contraction
by a factor  $e^{2N}$, where $N$ is the number of e-foldings of modes which
exit the Hubble radius during the
Ekpyrotic phase. The relative contributions of the
anisotropy and shear to the energy density decrease by this factor during the
phase of Ekpyrotic contraction.
In order to obtain a nonsingular bounce, $X$ must decrease after the end
of the Ekpyrotic phase by an even larger factor to reach the ghost condensate
phase with $X < X_c < X_i$. But this implies that the anisotropy and shear will increase
by the corresponding factor and will come to dominate again. A corollary
of this analysis is that the bouncing phase is long compared to the maximal
value of the Hubble expansion rate.

In our model, the ghost condensate is triggered not by $X$ decreasing, but
by $\phi$ increasing to some critical value. During the bouncing phase $X$
can remain large. Hence, the bounce phase will be short on the time scale
corresponding to the maximal value of the Hubble constant (this is
verified in our numerical results), and the anisotropy and shear will not increase.
Thus, it appears that our model is stable towards anisotropic stress
and shear instabilities.

Since the main idea of our model is to provide the adiabatic mode of
curvature fluctuations with a scale-invariant spectrum, the concern of
\cite{BingXue} regarding the spectrum of cosmological perturbations
does not arise. In fact, the mode which we are following is the same
mode as the one which becomes dominant after the bounce in \cite{BingXue}.
However, in our case it has a scale-invariant spectrum which is inherited
from the matter-dominated phase of contraction which preceded the
Ekpyrotic phase. We will return to this point at the end of the section on
cosmological perturbations.

\section{Cosmological Perturbations}

\begin{figure}
\includegraphics[scale=0.3]{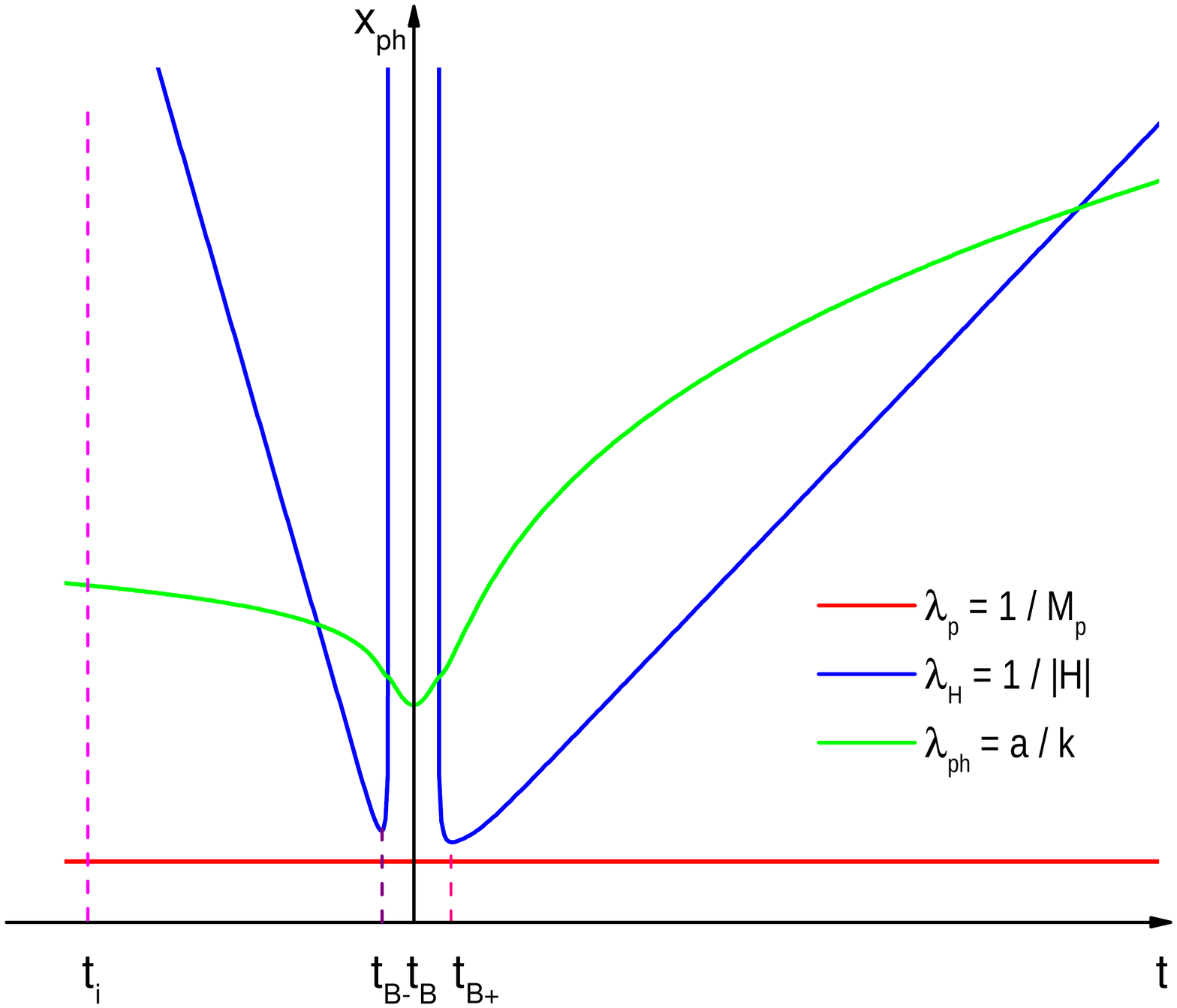}
\caption{A sketch of the evolution of length scales in the nonsingular bouncing
universe. The vertical axis is the physical spatial coordinate $x_{ph}$, and the
horizontal axis is the cosmic time $t$. The physical wavelength
$\lambda_{ph}=a/k$ of the mode with comoving wavenumber $k$ is depicted in green
(the second curve from the bottom at the earliest time); the Hubble radius
$\lambda_{H}=|H|^{-1}$ is depicted in blue (the curve which diverges to
infinity at the bounce point); and the red line (the constant bottommost line)
denotes the Planck length $\lambda_p=M_p^{-1}$. }
\label{Fig:sketchpert}
\end{figure}

We devote this section, to a study of the dynamics of linear cosmological perturbations in the
model. At the linearized level, each Fourier mode of the fluctuating field evolves independently.
It is useful to first consider a sketch of various relevant length scales in the
nonsingular bouncing cosmology. The first length scale is the physical wavelength
$\lambda_{ph}=a/k$ of the fluctuation mode (labelled by comoving wavenumber $k$)
which we wish to follow. This length must be compared with the
Hubble radius $H^{-1}(t)$. In order to allow for a causal generation mechanism
of fluctuations, the wavelength must be sub-Hubble at very early times. The third
length scale is the Planck length, the cutoff length below which our effective
field theory breaks down.  In Fig. \ref{Fig:sketchpert} we present a sketch of
the evolution of these length scales in the background cosmology.

As is obvious from Fig. \ref{Fig:sketchpert}, the perturbation modes we are
interested in originate inside the Hubble radius early in the contracting phase.
They exit the Hubble radius, briefly re-enter at the bounce point, and
finally enter the Hubble radius at late times in the period of Standard Big
Bang evolution. We will separate the time evolution of the fluctuations into
four phases: the initial era when the perturbations are set up (we will
consider several possibilities in this section), the evolution in the Ekpyrotic
contracting phase, the dynamics during the ghost condensate-dominated bouncing phase,
and the evolution in the post-bounce fast-roll expanding phase. We shall denote with
subscripts, ``$_{i}$", ``$_{B-}$", ``$_B$" and ``$_{B+}$" the initial moment when the
perturbations are set up, the beginning of the bouncing phase, the bounce point, and
the end of the bouncing phase, respectively. We will consider several choices
for initial conditions for the cosmological perturbation: vacuum fluctuations during
the Ekpyrotic phase, fluctuations formed in thermal equilibrium during the Ekpyrotic
phase, and the fluctuations inherited from a period of matter contraction (the case we
will be most interested in).

Since the equation of state is $w > - 1/3$, the  physical wavelength $\lambda_{ph}=a/k$
of a primordial fluctuation mode with fixed wavenumber $k$ decreases less fast
than the Hubble radius $\lambda_{H}=|H|^{-1}$. Thus, primordial perturbations
generated on sub-Hubble scales at the initial moment can escape into the
super-Hubble regime during the contracting phase. At late times in the
expanding phase, the Hubble radius catches up again and modes re-enter the
Hubble radius. Modes which we are interested in for cosmological observations
today re-entered the Hubble radius at very late times (at times close to or
after the time of equal matter and radiation). Thus, at the bounce point the
wavelength is many orders of magnitude larger than the minimal value of the
Hubble radius, which in turn is larger than the Planck length provided that the
scale of the bounce is sub-Planckian (which is a necessary requirement for
our effective field theory treatment to be justified). Thus, as in all
nonsingular bouncing cosmologies with bounce scale smaller than the Planck
mass, fluctuations never enter the sub-Planckian region of ignorance
(wavelengths smaller than the Planck length). This is a significant
advantage compared to the situation in inflationary cosmology where the
wavelengths of all fluctuation modes is smaller than the Planck length at
the beginning of the phase of inflation if inflation lasted more than
70 e-foldings \cite{TPproblem}. In our model, provided the energy scale
of the universe at the bounce is lower than the Planck scale, then the physical
wavelength of a perturbation mode corresponding to the current Hubble radius is
in the far infrared, as shown in Fig. \ref{Fig:sketchpert}. This lends
strong support to the use of linear cosmological perturbation theory to study
the evolution of the fluctuations.

\subsection{General Analysis}

Of the three families of metric fluctuations: scalar, vector and tensor modes,  we will concentrate
on scalar fluctuations, which couple at linear order in perturbation theory to matter. Vector perturbations
and tensor fluctuations (gravitational waves) do not couple to scalar field matter at linear order.
For matter given by a single scalar field there are - before
fixing coordinates - four scalar metric degrees of freedom and one matter degree of freedom. However, two of
these degrees of freedom can be eliminated by fixing coordinates. One degree of freedom is
constrained by the Einstein constraint equation, and yet another vanishes since there is no
anisotropic stress for scalar field matter. Hence, only one physical degree of freedom remains.
We choose this variable to be the curvature fluctuation $\zeta$ in uniform field gauge.

To obtain the action for scalar cosmological perturbations, we make a gauge fixed ansatz for
the metric and matter fluctuations, insert this ansatz into the Einstein equations and expand them
to leading order (in the amplitude of the fluctuations) about the cosmological background.
Another method of studying the dynamics of perturbations is to insert the ansatz for matter and
metric into the action for matter and gravity and expand to quadratic order in the fluctuations about
the cosmological background. Our detailed analysis is summarized in the first part of the
Appendix.  The resulting quadratic action for $\zeta$ is:
\begin{eqnarray}\label{S2_eta}
 S_2 = \int d\tau d^3x ~\frac{z^2}{2} \left[ \zeta'^2 - c_s^2(\partial_i{\zeta})^2 \right]~,
\end{eqnarray}
where we have introduced conformal time $\tau \equiv \int dt/a$.
The action is quadratic since we are working in linear perturbation theory. The
action contains no higher derivative terms  due to the special type of higher derivative
Lagrangian under consideration. Thus, the only change compared to what is obtained
in the case of a canonical scalar field coupled to Einstein gravity is the specific form of the
speed of sound $c_s$ and of the function $z$ which describes the coupling between the
background cosmology and the fluctuations.

The parameter $z$ is determined by the background metric and the background matter field.
Its form is derived in the Appendix and given by Eq. (\ref{z2_general}). The general expression
of the sound speed parameter $c_s$ is provided by Eq. (\ref{cs2_general}). In our
model, these two parameters take on the form
\begin{eqnarray}
\label{z2_model}
 z^2 &=& \frac{ 2M_p^4a^2\dot\phi^2 [M_p^2(1-g) +6\gamma{H}\dot\phi +3\beta\dot\phi^2 +\frac{3\gamma^2\dot\phi^4}{2M_p^2} ] }{(2M_p^2H-\gamma\dot\phi^3)^2}~, \nonumber\\
\label{cs2_model}
 c_s^2 &=& \frac{M_p^2(1-g) +4\gamma{H}\dot\phi +\beta\dot\phi^2 -\frac{\gamma^2\dot\phi^4}{2M_p^2} +2\gamma\ddot\phi}{M_p^2(1-g) +6\gamma{H}\dot\phi +3\beta\dot\phi^2 +\frac{3\gamma^2\dot\phi^4}{2M_p^2}}~.
\end{eqnarray}
It is easy to check that for $\beta = \gamma = g = 0$ the speed of sound becomes unity and
the form of $z$ reduces to the familiar one associated to a canonical scalar field in Einstein gravity
\cite{MFB, RHBrev}.

A common practice in the theory of cosmological perturbations, is to introduce
 a convenient quantity related to $\zeta$ by,
\begin{eqnarray}\label{vpara}
 v = z\zeta~,
\end{eqnarray}
which is a generalization of the well-known Mukhanov-Sasaki variable
\cite{Sasaki, Mukh}. In terms of this variable, the action (\ref{S2_eta}) takes on the
canonical form
\begin{eqnarray}
 S_2 = \int d\tau d^3x ~\frac{1}{2} \left[v'^2 -c_s^2(\partial_i v)^2 +\frac{z''}{z}v^2 \right]~,
\end{eqnarray}
where a prime indicates a derivative with respect to conformal time,
and thus lends itself to the process of canonical quantization. In Fourier space, the
equation of motion for the Fourier mode $v_k$ is
\begin{eqnarray} \label{flucteom}
 v_k'' + \left(c_s^2k^2-\frac{z''}{z} \right)v_k \, = \, 0~.
\end{eqnarray}

\subsection{Dynamics of Perturbations}

We discuss the solutions to the equation of motion for the cosmological fluctuations
in each of the phases of the background
evolution - the Ekpyrotic period, the bounce phase, and the kinetic-driven phase
of expansion after the bounce. On the transition hypersurface between the phases
we will continue the solutions making
use of the matching conditions derived by Hwang-Vishniac \cite{Hwang2}, and by
Deruelle-Mukhanov \cite{Deruelle}. These conditions indicate if the induced metric
on the matching surface and the extrinsic curvature are continuous. Note that we will
be using matching conditions at the end of the Ekpyrotic phase of contraction, and
at the end of the bounce phase. In both cases, the background dynamics also
obey these matching conditions. Thus, the difficulties of matching at a singular surface
between a contracting and an expanding phase which were discussed in \cite{Durrer}
are not present in our case. This procedure is the same as the one used in the past
(see e.g. \cite{Yifu1, Yifu2, Chunshan, HLbounce} where the results of the approximate
analytical calculations were also compared to direct numerical studies of the
fluctuation equations).

\subsubsection{Ekpyrotic contraction}

In the contracting phase, we have $|\phi| \gg1$ and $\dot\phi \ll M_p$. Thus one obtains
$g \simeq 0$, and in addition the higher order operators in the Lagrangian are suppressed.
In this case, we recover the same equations that apply for a normal canonical scalar field in
Einstein gravity. In the current case the scalar field potential is an exponential function. As
analyzed in the previous section, we obtain an Ekpyrotic contracting phase
and the background equation of state takes the value given in (\ref{eos_Ek}). In the
approximation of equality  in (\ref{eos_Ek}):
\begin{eqnarray}
 z^2 \simeq \frac{M_p^2a^2}{q}~,~~a \propto (\tilde\tau_{B-}-\tau)^{\frac{q}{1-q}}~,
\end{eqnarray}
and $c_s^2 \simeq 1$. We have introduced the time moment $\tilde\tau_{B-}$
when the scale factor would meet the big crunch singularity if there was no nonsingular bounce.
If we were not interested in the bouncing phase, it would make sense to normalize the time
axis such that $\tilde\tau_{B-} = 0$. In that case, we would find that
the function $g$ becomes unity slightly earlier, namely at a time $\frac{q}{(1-q){\cal H}_{B-}}$
(keeping in mind that ${\cal H}_{B-}$ is negative). This signals the beginning of the
bounce phase. In our case, we choose the time axis such that $\tau = 0$ is the midpoint
of the bounce phase and $\tau_{B-}$ is the beginning of this phase. In this case
\begin{eqnarray}
 \tilde\tau_{B-} = \tau_{B-}-\frac{q}{(1-q){\cal H}_{B-}}~.
\end{eqnarray}

Therefore, the equation of motion for cosmological perturbation in the contracting phase
simplifies
\begin{eqnarray}
 v_k'' + \bigg( k^2 -\frac{q(2q-1)}{(1-q)^2(\tau-\tilde\tau_{B-})^2} \bigg) v_k \simeq 0~,
\end{eqnarray}
and the general analytical solution is
\begin{eqnarray} \label{v_k^c}
 v_k^c(\tau) &=& c_1(k) \sqrt{\tau-\tilde\tau_{B-}} J_{\nu_c}[k(\tau-\tilde\tau_{B-})] \nonumber\\
 &&+ c_2(k) \sqrt{\tau-\tilde\tau_{B-}} Y_{\nu_c}[k(\tau-\tilde\tau_{B-})] ~,
\end{eqnarray}
with
\begin{eqnarray}\label{nu_c}
 \nu_c = \frac{1-3q}{2(1-q)} ~,
\end{eqnarray}
and the subscript ``$_c$" indicates that we are discussing the solution in the contracting
background as introduced in Sec. \ref{Sec:semi-analytic}. In Eq. (\ref{v_k^c}), $J$ and
$Y$ are two kinds of Bessel functions having indices  $\nu_c$. The coefficients
$c_1$ and $c_2$ are only functions of comoving wave number $k$, and they are determined
by the initial conditions of the cosmological perturbations which we will  address in a subsequent
subsection. For the moment, we keep the coefficients general.

Using the small argument limiting form of the Bessel functions, the first mode
in (\ref{v_k^c}) is decreasing in time on super-Hubble scales while the second mode is constant.
Since in the Ekpyrotic contracting phase the scale factor decreases only very slowly,
to first approximation, it remains true that the contribution to $\zeta$ from the second mode
dominates and  it is therefore approximately constant in time.

The initial power spectrum ${P}_{\zeta}$ (see Subsection D)
of the curvature fluctuations on super-Hubble scales can also be read off
from the small argument expansion of the Bessel functions: The dominant mode scales as
$Y_{\nu_c}(x)  \sim x^{-\nu_c}$ and hence the spectrum of the dominant mode of $\zeta$
is
\begin{equation} \label{inspec}
{P}_{\zeta} (k) \sim k^{3 - 2 \nu_v} |c_2(k)|^2 \, .
\end{equation}

\subsubsection{Nonsingular bounce}

When $\phi$ evolves into the ghost condensate range, the kinetic term is no longer
approximately canonical  and the Null Energy Condition is violated.
As we have discussed in the previous section, the
universe will exit from the phase of Ekpyrotic contraction at some moment $t_{B-}$, and
the equation of state of the universe will cross $w = -1$ and decrease
rapidly to negative infinity. Since in this period the $\dot\phi^2$ term in the expression
for the energy density (\ref{rho}) yields a negative contribution, it will eventually cancel
all the other positive contributions to the energy density. This happens at a moment $t_B$.
We choose our time axis such that $t_B = 0$. At this time, the Hubble parameter vanishes
and the nonsingular cosmological bounce occurs.

During the bounce phase, the deviation of the equation for fluctuations from the
canonical one becomes important.
When studying fluctuations in the bouncing phase, it has been shown in previous work
\cite{Yifu2, Yifu3, Chunshan} that it is a good approximation to model the evolution of the
Hubble parameter near the bounce as a linear function of cosmic time
\begin{eqnarray}\label{H_para}
 H = \Upsilon t~,
\end{eqnarray}
where $\Upsilon$ is a positive constant which has dimensions of $k^2$. This parametrization
is valid in a class of fast bounce models, and the magnitude of $\Upsilon$ is usually set by
the detailed microphysics of the bounce. In the specific example of our model with
background parameter values from Eq. (\ref{parameters}), $\Upsilon$ is of the
order $O(10^{-4})M_p^2$ (from Fig. \ref{Fig:Hubble}).

Next we consider the determination of the sound speed square in the neighborhood of the
bounce. From the expression (\ref{cs2_model}), we find that, in addition to the Hubble
parameter, one needs to know the form of $\dot\phi^2$ around the bounce. We have
evaluated this in our semi-analytical study of the background solution and the result
is given in Eq. (\ref{dotphiB}). Combining Eqs. (\ref{dotphiB}), (\ref{cs2_model}) and
(\ref{H_para}), we find that the sound speed parameter takes the approximate form
\begin{eqnarray}\label{cs2_b}
 {c_s^2}_b \simeq \frac{1}{3}-\frac{2}{3\sqrt{1+\frac{12\beta V_0}{M_p^4(g_0-1)^2}}}~,
\end{eqnarray}
in the bouncing phase. The subscript ``$_b$" in Eq. (\ref{cs2_b}) indicates the bouncing
phase. From this result, we see that the perturbations in the model are not completely stable
around the bounce since it is possible for $c_s^2$ to be negative during the bouncing
phase. For example, if we make use of the parameter choice (\ref{parameters}), we
obtain ${c_s^2}_b \simeq -\frac{1}{3}$. A negative sound speed square leads to an
exponential growth of the perturbation modes. We will, however, now show that
this growth is not too large to loose perturbative control of the analysis.

To calculate the growth of perturbations during the bounce phase, we return to the
equation of motion (\ref{flucteom}) for the fluctuation modes.
In that equation, the key quantity is the parameter $z$ whose explicit form is given by Eq.
(\ref{z2_model}). In our model, $\dot\phi$ takes on its maximal value at the bounce point, as
has been analyzed in Section II-D. The explicit expression is given in Eq. (\ref{dotphiB}).
When we choose the parameter $V_0$ to be very small, we obtain the approximate value
\begin{eqnarray}
 \dot\phi^2_{B} \, \simeq \, \frac{2M_p^2(g_0-1)}{3\beta}~,
\end{eqnarray}
and thus we can simplify the expression for $z^2$:
\begin{eqnarray}
 z^2 \, \simeq \, a^2 \frac{3\beta M_p^4\dot\phi^4}{(2M_p^2H-\gamma\dot\phi^3)^2}~,
\end{eqnarray}
in the bouncing phase.

The dynamics of the $z$ parameter during the bouncing phase depend mainly on the
Hubble parameter $H$ and the time derivative of the scalar $\dot\phi$. We find that
the evolution of $\dot\phi$ is approximated by
\begin{eqnarray}
 \dot\phi \, \simeq \, \dot\phi_B e^{-\frac{t^2}{T^2}}~,
\end{eqnarray}
where $T$ is approximately one quarter of the duration of the bounce (as shown in Fig. \ref{Fig:Graph_dotphi}). In the following figure we compare this
approximate expression with the numerical result for $\dot\phi$.

\begin{figure}
\includegraphics[scale=0.3]{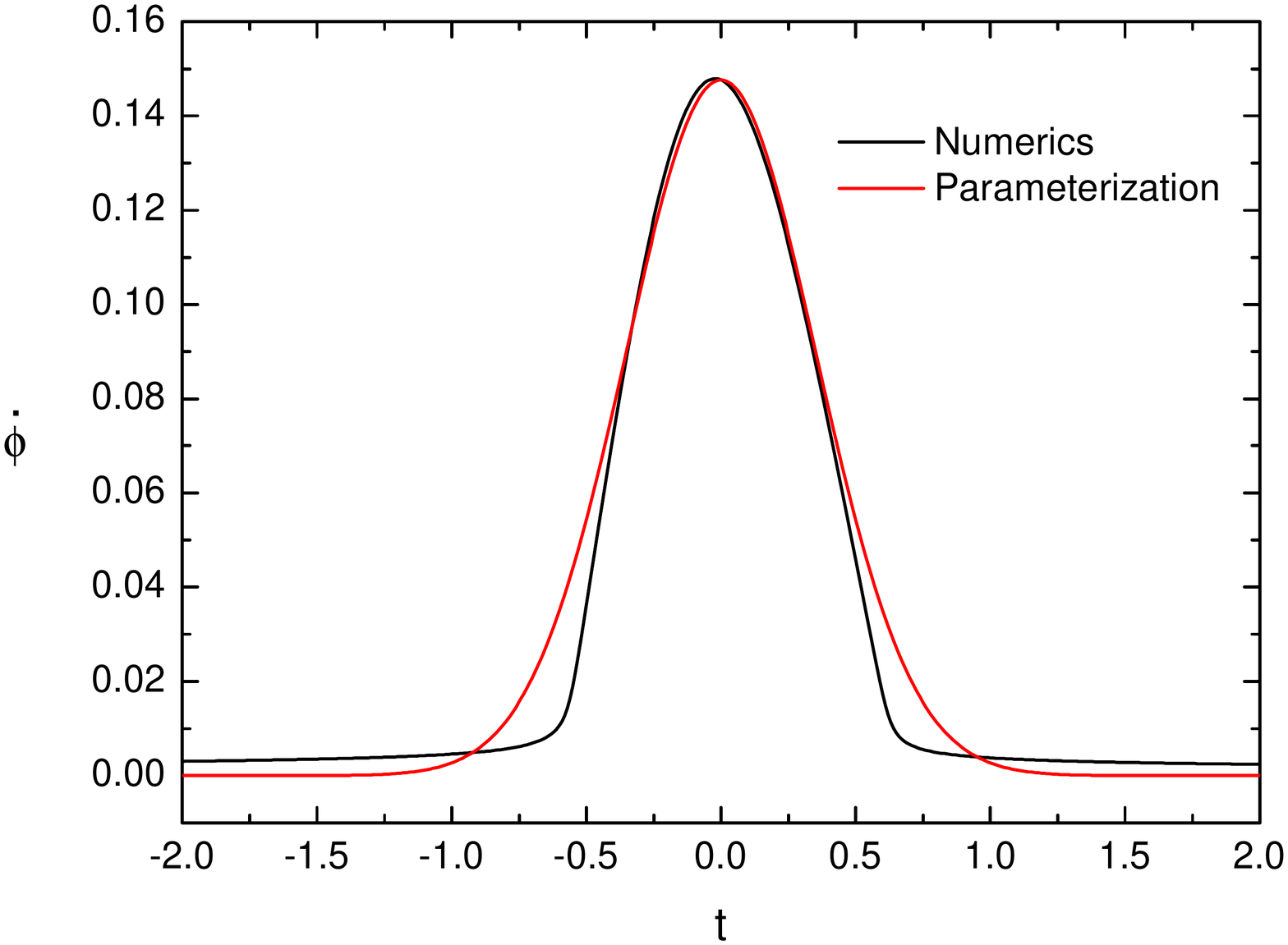}
\caption{Numerical result and analytical parameterization of $\dot\phi$ as a function of
cosmic time around the bounce point. }
\label{Fig:Graph_dotphi}
\end{figure}

Inserting the above parameterizations for $\dot\phi$ and (\ref{H_para}) for $H$
into the expression of $z^2$,  we can neglect the contribution from $H$
and further simplify the result
\be
z^2\, \simeq \, a^2\frac{3\beta M_p^4}{\gamma^2\dot\phi^2} \, .
\ee
Therefore, we obtain
\begin{eqnarray}
 \frac{z''}{z} \, \equiv \, a^2 M^2
\end{eqnarray}
with
\begin{eqnarray}
 M^2 \, \simeq \, \Upsilon + \frac{2}{T^2} +
(2\Upsilon^2 + \frac{6\Upsilon}{T^2} + \frac{4}{T^4})t^2~.
\end{eqnarray}
Consequently, the perturbation equation (\ref{flucteom}) can be
approximated as follows:
\begin{eqnarray}
 v_k'' - \omega^2v_k \, \simeq \, 0~,
\end{eqnarray}
with
\begin{eqnarray} \label{omegaeq}
 \omega^2 \, = \, -{c_s^2}_b k^2 + a_B^2 M^2~,
\end{eqnarray}
in the bouncing phase. Here, $a_B$ is the value of
the scale factor at the bounce point which we set
to unity. One of the solutions is exponentially growing, the
other one decaying:
\begin{eqnarray}\label{v_k^b}
 v_k^b(\tau) \, \simeq \, d_1(k) e^{\omega(\tau-\tau_B)} + d_2 e^{-\omega(\tau-\tau_B)}~,
\end{eqnarray}
where $d_1$ and $d_2$ are two coefficients which can be determined by the matching
conditions for fluctuations, and both of which inherit the spectrum of the dominant mode
in the contracting phase.

Note that for the infrared modes which we are interested in the first term in (\ref{omegaeq})
is negligible. Hence, the growth rate of infrared modes during the bounce phase is
independent of $k$. The amplification factor ${\cal F}$ for the perturbation modes during the
bounce phase is
\begin{eqnarray}
 {\cal F} \, &=& \, e^{\int_{B-}^{B+}\omega d\tau}  \nonumber\\
 &\simeq& \exp \left[ \sqrt{\Upsilon+\frac{2}{T^2}}t  + \frac{2+3\Upsilon T^2+\Upsilon^2 T^4}{3T^4\sqrt{\Upsilon+\frac{2}{T^2}}} t^3 \right] \bigg|_{B-}^{B+}~. \nonumber\\
\end{eqnarray}
Inserting the values of the parameters $\Upsilon=2.7\times10^{-4}$, $T=0.5$, and
$t_{B+}=-t_{B-}=1$ (obtained from the numerical solution for the cosmological
background) we obtain
\begin{eqnarray}
 {\cal F} \, \simeq \, e^{11} \, \sim \, O(10^5)
\end{eqnarray}
which is in approximate agreement with our numerical calculation as we demonstrate
shortly. Note that this growth rate is the same for all
infrared modes of interest for current cosmological observations.

\subsubsection{Fast-roll expansion}

After the bounce, the $\dot\phi^2$ term will decrease back to a value below the
Planck scale and the scalar field $\phi$ will cross the critical value $\phi_{*+}$. Therefore,
the phase of ghost condensation will cease at a moment $t_{B+}$ after the bounce and
the universe will smoothly enter an expanding era.  Afterwards, all higher order operators
are suppressed by the Planck scale and thus the Lagrangian
recovers its canonical form. Since the potential is very flat in the expanding era,
the scalar field will, as discussed in the previous section, enter a fast-roll state with
equation of state $w \simeq 1$. In this period, the equation of motion for cosmological
perturbations is given by
\begin{eqnarray}
 v_k''+ \left(k^2+\frac{1}{4(\tau-\tilde\tau_{B+})^2} \right)v_k \, \simeq \, 0~,
\end{eqnarray}
and yields the following solution
\begin{eqnarray}\label{v_k^e}
 v_k^e(\tau) &\simeq& e_1(k) \sqrt{\tau-\tilde\tau_{B+}} J_0[k(\tau-\tilde\tau_{B+})] \nonumber\\
 &&+ e_2(k) \sqrt{\tau-\tilde\tau_{B+}} Y_0[k(\tau-\tilde\tau_{B+})]~,
\end{eqnarray}
with $\tilde\tau_{B+}\equiv\tau_{B+}-\frac{1}{2{\cal H}_{B+}}$.
The superscript ``e" indicates that we are talking about the solution in the expanding phase.

Modulo the square root factor, the first mode is constant whereas the second one is
increasing logarithmically on super-Hubble scales. Hence, the second mode dominates at late times.
In particular, we will be interested in the spectrum of that mode.

\subsection{Matching Conditions}

Having obtained general solutions to the perturbation equations in the various phases of
the cosmological evolution, we now study how the solutions are to be matched at the
transition points between the phases, i.e. we must determine the coefficients appearing in
the solutions (\ref{v_k^b}) and (\ref{v_k^e}) through the matching conditions. As
mentioned at the beginning of this subsection, the matching conditions tell us that the
induced 3-metric on the hyper-surface of the phase transition and its extrinsic curvature
should be continuous across the matching surface. In an expanding universe, it
has been shown \cite{BST, BK} that in the absence of entropy fluctuations
the curvature perturbation $\zeta$ in constant field gauge
is a conserved quantity on large scales (this result extends even beyond linear
perturbation theory \cite{Langlois}), and thus $\zeta$ is the variable which is now generally
used to describe the dynamics of cosmological perturbations. However, in a contracting
background there are complications: $\zeta$ has a growing mode, and care needs to
be taken when applying the matching conditions. A second reason which forces us to
reconsider the matching condition issue is that, when the universe experiences moments
of violating the Null Energy Condition, i.e., $w$ crossing $w = -1$, the variable $\zeta$
may not be well behaved since there are factors of $(1+w)$ which in the standard theory
relate $\zeta$ to other metric fluctuation variables. Therefore, we must consider the
detailed dynamical evolution of the metric fluctuations in each phase and carefully study
the matching conditions.

In our model, we have chosen the uniform field gauge (see the first part of the Appendix).
Since the field $\phi$ is monotonically increasing in our scenario, this gauge is well-defined
throughout, unlike what happens during inflationary reheating at a turning point of $\phi(t)$.
Our matching surfaces are the $\phi = \phi_{*-}$ and $\phi = \phi_{*+}$ surfaces. The
matching conditions for the background are satisfied on both matching surfaces (unlike
what would happen if we were to try to match between a contracting branch and an
expanding branch on a singular transition surface). Hence, the
matching conditions of \cite{Hwang2, Deruelle} apply, indicating that $v$ and $v'$ are
continuous across the matching surfaces, i.e. at the moments $t_{B-}$ and $t_{B+}$.
Recall that both solutions of the perturbation equations on super-Hubble scales in
both the contracting phase and kinetic-driven expanding phase are Bessel
functions. We can use their asymptotic forms to simplify the calculations.

At the first transition surface, the transition between the contracting
phase and the nonsingular bounce period, it is sufficient to consider the asymptotic
form of the solution (\ref{v_k^c}) in the contracting phase:
\begin{eqnarray}\label{v_k^c_asym}
 v_k^c(\tau) &\simeq& \frac{c_1(k) k^{\nu_c} }{2^{\nu_c} \Gamma_{1+\nu_c}} (\tau-\tilde\tau_{B-})^{\frac{3w_c-1}{3w_c+1}} \nonumber\\
 && - \frac{2^{\nu_c} \Gamma_{\nu_c} c_2(k) }{\pi k^{\nu_c}} (\tau-\tilde\tau_{B-})^{\frac{2}{3w_c+1}}~,
\end{eqnarray}
where $\Gamma_{\nu_c}$ is the Gamma function of $\nu_c$-th order. To obtain this formula,
we have made use of the expression for the index $\nu_c$ (\ref{nu_c}) and of the inverse of
the relation of Eq. (\ref{eos_Ek}) to derive the explicit evolution of the fluctuation on
super-Hubble scales as a function of conformal time. If the coefficients $c_1$ and $c_2$
are of the same order (whether this is the case or not is determined by the initial
conditions, as we shall discuss below) the second mode of $v_k^c$ (with
coefficient function $c_2(k)$) dominates at late stages of the contraction phase
since the $c_1$ term decreases faster as a function of time if $w_c$ is greater than $1$.
\footnote{It is worth while to mention that the dominant mode is the $c_1$ term for a
bouncing cosmology with $w_c$ less than $1$, e.g. in the ``Matter Bounce" \cite{RHB2011rev}
cosmology. Since in general the cosmological perturbations after the bounce will have a spectrum
which is determined mainly by the dominant mode of the fluctuations in the contracting phase,
a different dominant mode will give rise to different results after applying the matching
conditions.}

By requiring $v_k$ and $v_k'$ to be continuous at the space-like surfaces of
$\tau_{B-}$ and $\tau_{B+}$, one can track how each Fourier mode evolves through
the bouncing phase and derive the detailed expressions of all the coefficients appearing
in the above solutions of the perturbations equations. We leave the detailed calculation
for the second part of the Appendix, and here just write down the final expression of the
perturbation which is
\begin{eqnarray}\label{v_k^e_asym}
 v_k^e(\tau) \, \simeq \, {\cal F} \frac{ \gamma_E \Gamma_{\nu_c} c_2(k) k^{-\nu_c} }{ 2^{1-\nu_c} \pi (\tau_{B-}-\tilde\tau_{B-})^{\nu_c-\frac{1}{2}} } \bigg(\frac{\tau-\tilde\tau_{B+}}{\tau_{B+}-\tilde\tau_{B+}}\bigg)^{\frac{1}{2}}~,
\end{eqnarray}
where $\gamma_E \simeq 0.58$ is the Euler-Masheroni constant.

Let us comment on the time dependence and spectrum of the resulting cosmological fluctuations, first
focusing on the amplitude. While the universe expands with an equation of state $w = 1$, the
scale factor evolves as $a \sim \tau^{\frac{1}{2}}$. This implies that the perturbation variable
$v_k^e$ evolves proportional to the scale factor and therefore the curvature perturbation
$\zeta$ will become conserved on super-Hubble scales after the bounce, as it must since
there are no entropy fluctuations in the model.

The spectrum shape is set by the spectrum of the coefficient function $c_2(k)$ from the contracting
phase. The late time power spectrum ${P}_{\zeta}$ (see following subsection for a
more detailed discussion) of the curvature fluctuation variable $\zeta$,
the power spectrum relevant for late time observations, is proportional to the phase space factor
$k^3$ multiplied by the square of the absolute value of the mode function, i.e.
\begin{equation} \label{finspec}
{P}_{\zeta}(k) \, \sim k^{3 - 2 \nu_c} |c_2(k)|^2 \, .
\end{equation}
Comparing with the spectrum of $\zeta$ in the contracting phase (\ref{inspec}) we see that
the spectrum of $\zeta$ has passed through the nonsingular bounce without change in
its spectral index, but with boosted amplitude. This is the main result of this section.

\subsection{Initial conditions}

In this subsection, we investigate a group of initial conditions which set the momentum
dependence of the coefficient $c_2(k)$. We determine the conditions necessary
 to obtain a scale-invariant primordial power spectrum.  Recall
the definition of the power spectrum of primordial curvature perturbations
\begin{eqnarray}\label{P_zeta}
 P_{\zeta} &\equiv& \frac{k^3}{2\pi^2} |\zeta_k|^2 = \frac{k^3}{2\pi^2} |\frac{v_k}{z}|^2 \\
 &=& \frac{k^3}{6\pi^2M_p^2} |\frac{v_k^e}{a}|^2~.
\end{eqnarray}
The spectral index is defined as
\begin{eqnarray}\label{n_s}
 n_s-1 \equiv \frac{d\ln P_{\zeta}}{d\ln k}~.
\end{eqnarray}

Inserting the solution (\ref{v_k^e_asym}) into Eq. (\ref{P_zeta}) and making use of the
index (\ref{nu_c}), we see that the primordial power spectrum will be scale-invariant
(in the case $w_c > 1$) when
\begin{eqnarray}\label{SI_condition}
 c_2(k)  \sim k^{\nu_c-\frac{3}{2}} \sim k^{-\frac{3(1+w_c)}{1+3w_c}} ~,
\end{eqnarray}
is satisfied\footnote{In the case of $w_c<1$, e.g. as in the original ``Matter Bounce"
cosmology, the $c_1$ mode dominates during the contracting phase. As a consequence,
the matching conditions are modified and the condition for scale invariance becomes
$c_1(k) \sim k^{-\nu_c-\frac{3}{2}} \sim k^{-\frac{6w_c}{1+3w_c}}$. }.

We will first consider perturbations originating as vacuum fluctuations in the phase
of Ekpyrotic contraction, then perturbations which originate as thermal particle
fluctuations, and finally the main case studied here, initial conditions inherited
from vacuum perturbations which exit the Hubble radius in a matter-dominated
phase of contraction which preceded the Ekpyrotic phase.

\subsubsection{Ekpyrotic vacuum fluctuations}

First, we consider cosmological perturbations which originate as quantum
vacuum fluctuations on sub-Hubble scales in the Ekpyrotic phase. Since the action
for $v_k$ is that of a harmonic oscillator, quantum vacuum initial conditions mean
\begin{eqnarray}\label{v_k^i_vacuum}
 v_k(\tau) \rightarrow \frac{e^{-ik(\tau-\tilde\tau_{B-})}}{\sqrt{2k}}~.
\end{eqnarray}
To match this initial condition with the sub-Hubble perturbations in contracting
phase (\ref{v_k^c}), we get
\begin{equation}
c_1 \sim c_2 \sim k^0 \, .
\end{equation}
Explicitly, we obtain
\begin{equation}
c_2 \simeq -\frac{\pi}{2^{\nu_c+\frac{1}{2}}\Gamma_{\nu_c}} \, ,
\end{equation}
Inserting this result into the equation for the modes in the expanding phase, we obtain
the final expression
\begin{eqnarray}\label{v_k^e_vacuum}
 v_k^e(\tau) \simeq  -\frac{ {\cal F} \gamma_E a(\tau) }{ 2^{\frac{3}{2}} k^{\nu_c} (\tau_{B-}-\tilde\tau_{B-})^{\nu_c-\frac{1}{2}} a_{B+} }~.
\end{eqnarray}

Therefore, for vacuum initial conditions in the contracting phase, the power spectrum of
primordial cosmological perturbations becomes
\begin{eqnarray}\label{P_zeta_vacuum}
 P_\zeta \, \simeq \,
{\cal F}^2 \frac{2^{\frac{4}{1+3w_c}}\gamma_E^2H_{B-}^2}{48\pi^2 (1+3w_c)^\frac{4}{1+3w_c} M_p^2} \left(\frac{k}{|{\cal_H}_{B-}|}\right)^{\frac{6(1+w_c)}{1+3w_c}}~.
\end{eqnarray}
From this result, we can read that the largest amplitude of power spectrum is of order
${\cal F}^2(\frac{H_{B-}}{M_p})^2$, the vacuum amplitude of the perturbation mode which exited
the Hubble radius at the time $t_{B-}$ and for which $k\sim |{\cal H}_{B-}|$, multiplied
by the boost factor ${\cal F}^2$. Modes with $k>|{\cal H}_{B-}|$ will never exit the Hubble radius, they never undergo squeezing on super-Hubble scales, and hence remain as vacuum quantum fluctuations.
Combining the above result (\ref{P_zeta_vacuum}) with the definition of the spectral index
from Eq. (\ref{n_s}), we find that the spectral index for observable perturbation modes is
\begin{eqnarray} \label{ekpspectrum}
 n_s = \frac{7+9w_c}{1+3w_c}~,
\end{eqnarray}
which is always blue tilted when $w_c$ is bigger than $1$ and converges to $3$ when
$w_c$ is large. This result agrees with previous studies of artificially smoothed out
four-dimensional toy models of Ekpyrotic cosmology \cite{Lyth, Fabio2, Tsujikawa}
\footnote{Note that the fluctuations in Ekpyrotic models motivated from string theory may
well be scale-invariant since the extra spatial dimensions \cite{Thorsten}, entropy
modes \cite{Notari, Finelli, Creminelli, Turok, Khoury} and stringy effects (see e.g. \cite{Tolley})
play an important role.}.

\subsubsection{Ekpyrotic thermal fluctuations}

In a bouncing cosmology, it is not manifest that primordial perturbations have to arise
as quantum vacuum fluctuations. An alternative choice for conditions for the primordial
perturbations is thermal fluctuations of a gas of particles. If the fluctuations are
generated by a thermal ensemble of point particles, e.g. relativistic particles with
equation of state $w_r = 1/3$, then it was shown that an Ekpyrotic contracting phase
with a particular background equation of state, namely $w_c \simeq 7/3$ is required to
produce a nearly scale-invariant power spectrum \cite{Cai:2009rd}. In the following we will
revisit this issue in the context of the explicit model of this paper. Our result is in
agreement with the general analysis performed in Ref. \cite{Cai:2009rd}. Moreover, we
obtain the explicit relation between the spectral index and the model parameters.

Consider that the universe is near thermal equilibrium at the beginning of the Ekpyrotic
contracting phase. In that case, thermal fluctuations will be more important than
vacuum perturbations. In a thermal system, the mean square
mass fluctuation in a sphere of radius $R$ (and volume $V$)
is given by the specific heat capacity $C_v(R)$
\begin{eqnarray}\label{M2}
 \langle\delta{M}^2\rangle = C_v(R) T^2~,
\end{eqnarray}
where $T$ is the temperature of the thermal system. For a gas of normal particles,
the heat capacity is extensive:
\begin{eqnarray}\label{Cv}
 C_v = R^3\frac {\partial\rho}{\partial T} \sim R^3T^3~.
\end{eqnarray}
The mean square mass fluctuations in a sphere of radius $R$ are given in terms of the
k'th momentum mode (with $k  = a R^{-1}$) multiplied by the phase space factor $k^3$
in the following way (see \cite{RHBrev})
\begin{eqnarray}\label{rho2}
 \langle \delta{M}^2 \rangle \simeq V^2 k^3 \langle \delta\rho_k^2 \rangle~,
\end{eqnarray}
where $V = \frac{4 \pi}{3} R^3$. Inserting (\ref{M2}) and (\ref{Cv}) into (\ref{rho2}), one obtains
\begin{eqnarray}
 \delta\rho_k \simeq O(1) k^{-\frac{3}{2}} R^{-\frac{3}{2}} T^{\frac{5}{2}}~.
\end{eqnarray}

The density fluctuations determine the gravitational potential $\Phi$
through the time-time component of the perturbed Einstein equations
(which is also the so-called relativistic Poisson equation), and this
in turn allows us to compute the Mukhanov-Sasaki variable $v_k$.
The relations are
\begin{eqnarray}
 \frac{k^2}{a^2} \phi_k^{i} &\simeq& 4\pi G \delta\rho_k ~,  \\
 \Phi_k &=& -\frac{a^2 \dot{H}}{k^2H} \left(\frac{M_p v_k}{z_k}\right)^{.}~,
\end{eqnarray}
while the length scale of the perturbation is smaller than the Hubble radius.
As a consequence (making use of the fact that $v_k$ oscillates in time with
frequency $k/a$ on sub-Hubble scales), the amplitude of the Mukhanov-Sasaki variable at
Hubble crossing moment can be expressed as
\begin{eqnarray}
 v_k(\tau_H) \simeq \frac{aHz_k\delta\rho_k}{2 i M_p^3 k \dot{H}}|_{R=\frac{1}{H}}
 \simeq \frac{i a_k^{\frac{3}{2}} T_k^{\frac{5}{2}}}{2 \epsilon_H M_p^2 k^2}~,
\end{eqnarray}
where we have applied the Hubble radius crossing condition $k=aH$.

Since the temperature of the universe scales as $T\sim a^{-1}$
for normal radiation and since in the Ekpyrotic contracting phase
$a\sim\tau^{\frac{q}{1-q}}$, and since $\tau_H\sim 1/k$  is
the Hubble crossing condition, we can derive the $k$
dependence of the perturbation variable. The result is
\begin{eqnarray}
 v_k(\tau_H(k)) \sim k^{-2+\frac{q}{1-q}}~,
\end{eqnarray}
which implies
\begin{eqnarray}\label{c2_themal}
 c_2(k) \simeq \frac{(-\pi)}{2^\nu_c\Gamma_{\nu_c}}k^{\frac{1}{2}} v_k(\tau_H(k))
 \sim k^{\frac{1-9w_c}{2(1+3 w_c)}}~.
\end{eqnarray}
Comparing Eq. (\ref{c2_themal}) and the condition of scale invariance (\ref{SI_condition}),
we conclude that only when $w_c \simeq \frac{7}{3}$ the primordial power spectrum
of curvature perturbations seeded by thermal fluctuations  will be scale-invariant.
This agrees with the result of \cite{Cai:2009rd}.

\begin{figure}
\includegraphics[scale=0.3]{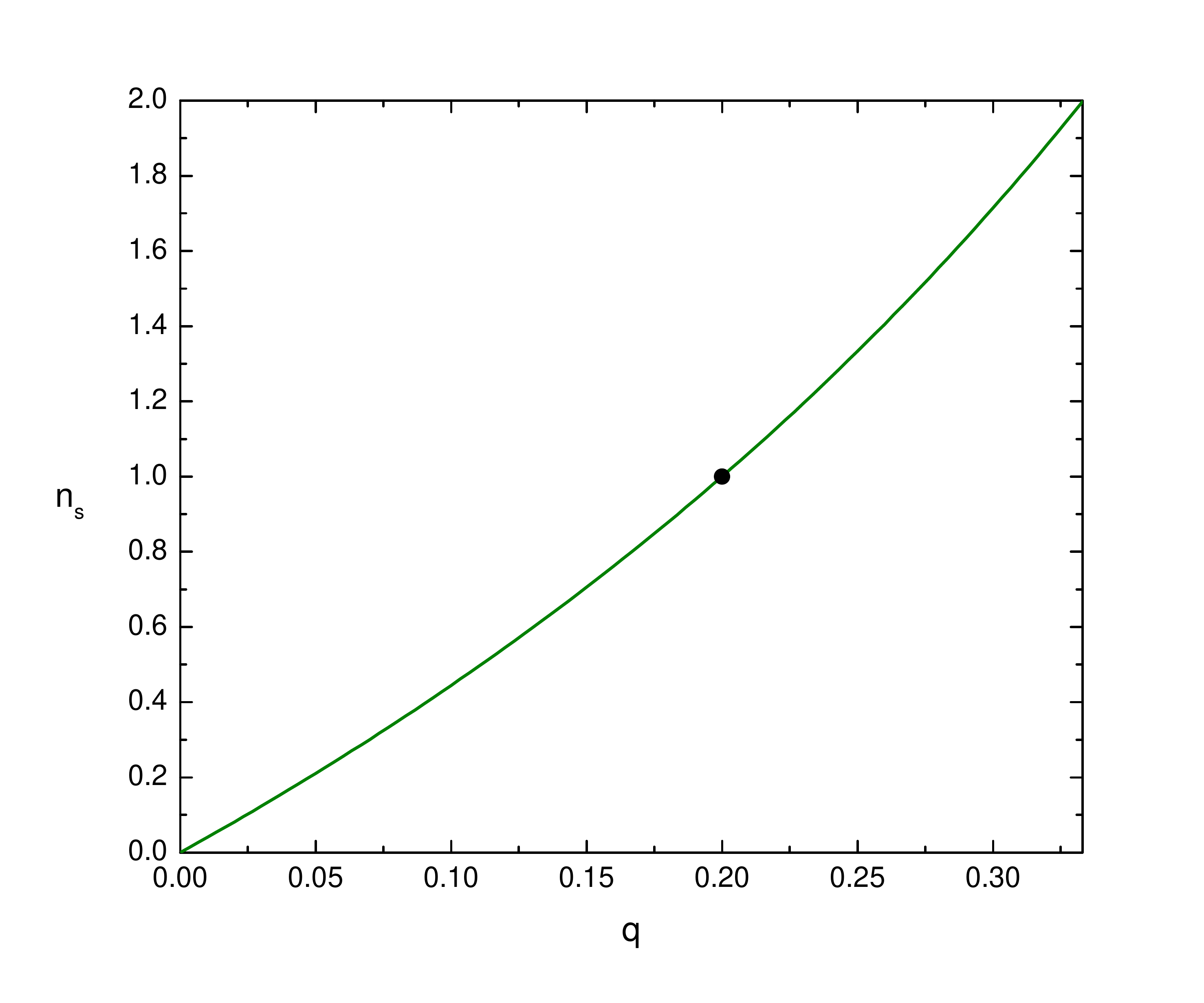}
\caption{The spectral index of primordial power spectrum $n_s$ (vertical axis)
as a function of the parameter $q$ (horizontal axis). The
black dot corresponds to the case of exact scale invariance of the spectrum
when $q=0.2$.}
\label{Fig:ns}
\end{figure}

Explicitly, the spectral index of the primordial power spectrum is determined by
$q$ as follows:
\begin{eqnarray}
 n_s = \frac{4q}{1-q}~.
\end{eqnarray}
We plot the spectral index of the primordial power spectrum seeded by thermal fluctuations
as a function of the model parameter $q$ in Fig. \ref{Fig:ns}. From this figure, one can read
off that when $q$ is slightly less than $0.2$, the primordial power spectrum would have a
slight red tilt, a result which is favored by current CMB observations.

To summarize,  we have shown that, assuming thermal initial condition for cosmological
perturbations in an Ekpyrotic phase of contraction, our model is able to generate a nearly
scale-invariant power spectrum by choosing a suitable parameter $q$.

\subsubsection{Fluctuations inherited from a phase of matter-dominated contraction}

If there is no matter and entropy generation during the bounce phase
\footnote{If the cosmological fluctuations remain in the linear regime, then one
can show that no entropy is produced \cite{Tomislav}.}, then the universe
must have originated from a contracting phase which in the far past was dominated by
matter and radiation. In particular, scales which we probe today in cosmological observations
are likely to have exited the Hubble radius during a matter-dominated contracting phase,
and those which are probed by smaller scale observations (e.g. QSO observations) would have
been exited during a radiation-dominated contracting phase. Thus, it is logical to consider
the spectrum of fluctuations in this context. Specifically, we must
compute the spectrum of fluctuations for modes which have exited
the Hubble radius in a pre-Ekpyrotic phase of contration. We will focus exclusively
on adiabatic perturbations (given that we now have several matter fields, namely regular
cold matter, radiation plus the scalar field $\phi$, there is the possibility of entropy
fluctuations).

To compute the spectrum of perturbations we make use of the gauge-invariant variable
$\zeta$, the curvature fluctuation in comoving coordinates, and the corresponding
Mukhanov-Sasaki variable $v = z \zeta$, where $z\equiv \sqrt{2\epsilon}a$ with
$\epsilon\equiv -\dot H/H^2$. The equation of motion for the Fourier mode $v_k(\tau)$
in the context of standard Einstein gravity is
\begin{eqnarray}\label{eom_matter}
 v_k''+(k^2-\frac{z''}{z})v_k = 0~,
\end{eqnarray}
where the prime denotes the derivative with respect to the comoving time as defined
in the previous section. For a constant background equation of state $w_m=0$, one obtains
\begin{eqnarray}\label{mass}
 \frac{z''}{z} = \frac{\nu_m^2-\frac{1}{4}}{\tau^2}~,~
 {\rm with}~~\nu_m=\pm\frac{3}{2}~.
\end{eqnarray}

We assume that the cosmological perturbations during the matter-dominated period
of contraction originate from vacuum fluctuations, which implies that on
sub-Hubble scales we must have
\begin{eqnarray}\label{inicond}
 v_k^{i}\simeq\frac{1}{\sqrt{2k}}e^{-i\int^\tau kd\tilde\tau}~,
\end{eqnarray}
(for $|k\tau|\gg1$). This is consistent with the vacuum initial condition provided in
Eq. (\ref{v_k^i_vacuum}) when the last term $\frac{z''}{z}$ is negligible.

On super-Hubble scales we can use the small argument expansion of the Bessel function
solution of Eq. (\ref{eom_matter}) to find the asymptotic form
\begin{eqnarray}\label{sollead}
 v_k^m \sim \tau^{\frac{1}{2}} \bigg[ c_m(k)\tau^{-\frac{3}{2}} \bigg]~,
\end{eqnarray}
(for $|k\tau|\ll1$).

Matching the two asymptotic solutions (\ref{inicond}) and (\ref{sollead}) at the moment
of Hubble crossing $|k\tau|\sim1$ yields the final form of the solution for $v_k^m$
on super-Hubble scales
\begin{eqnarray}\label{v_k^m}
 v_k^m(\tau) \simeq \frac{1}{\sqrt{2k^3}(\tau-\tilde\tau_m)}~,
\end{eqnarray}
where $\tilde\tau_m = \tau_m - 2/{\cal H}_m$ with $\tau_m$ defined as the end point
of the period of matter contraction, and ${\cal H}_m$ is the comoving Hubble parameter at
that time. Thus, we have reproduced the well known result that the spectrum of curvature
fluctuations originating from vacuum perturbations on scales which exit the Hubble radius
during a matter-dominated phase of contraction is scale-invariant \cite{Wands, Fabio2}
(see also the detailed calculations in the Lee-Wick bounce model \cite{Cai:2008qw}).

Since we have shown that the spectrum of curvature perturbations passes through
the bounce phase with unchanged index, we conclude that on the scales considered here,
the final power spectrum of curvature fluctuations will be scale-invariant. Thus, we have shown
that our current model provides a realization of the ``matter bounce" scenario.

To see this result in a more detail,
we need to match the initial condition (\ref{v_k^m}) with the asymptotical solution of
$v_k$ in the Ekpyrotic contracting phase (\ref{v_k^c_asym}) at the moment
$\tau_m$. This gives us the form of $c_2(k)$:
\begin{eqnarray}
 c_2(k) \simeq \frac{ (-\pi) k^{\nu_c-\frac{3}{2}} (\tau_m-\tilde\tau_{B-})^{\nu_c-\frac{1}{2}} }{ 2^{\nu_c+\frac{1}{2}} \Gamma_{\nu_c} (\tau_m-\tilde\tau_m) }~,
\end{eqnarray}
and thus we can derive the expression of the canonical perturbation variable in the
expanding phase of our model, the result being
\begin{eqnarray}\label{v_k^e_matter}
 v_k^e(\tau) \simeq -{\cal F}\frac{\gamma_EH_m}{2^{\frac{5}{2}}k^{\frac{3}{2}}} a(\tau)~,
\end{eqnarray}
where we have made use of Eq. (\ref{v_k^e_asym}).

As a consequence, the scale invariance of primordial fluctuations which exited  the Hubble
radius during the matter-dominated phase of contraction is preserved through the
Ekpyrotic phase and the nonsingular bouncing phase, and the final power spectrum will
be scale-invariant in the expanding phase. To find the amplitude of the spectrum, we
apply the definition of $P_\zeta$ to obtain
\begin{eqnarray}\label{amplitude}
 P_{\zeta} \, \simeq \, {\cal F}^2 \frac{\gamma_E^2 H_m^2}{192\pi^2M_p^2}~,
\end{eqnarray}
where the parameter $H_m$ is the physical Hubble parameter at the end of matter-dominated
period of contraction.

Notice that if there is regular matter and radiation which dominated early in the contracting
phase, the overall spectrum of fluctuations will change its shape. It will be scale-invariant on
very large scales (those which exited the Hubble radius during matter contraction), it will
then shift to being a vacuum spectrum for scales which exited in the radiation phase of
contraction \cite{Hongli}, and will end up with the deep blue spectrum of (\ref{ekpspectrum})
on scales which exit during the period of Ekpyrotic contraction.

\subsection{Numerical Analysis of Cosmological Perturbations }

Previously, we gave an analytic estimate of the amplitude of primordial perturbations
generated for various initial conditions in our model. The analysis in the bounce phase
was rather non-trivial and the fluctuation modes undergo a period of exponential
growth. In order to confirm the analytical approximations made and to check that
the model is indeed well-behaved when passing through the
nonsingular bounce point at the perturbative level, we numerically studied the evolution
of primordial perturbations throughout the bounce. Specifically, we considered the case
of vacuum initial condition generated in the Ekpyrotic phase of contraction.

\begin{figure}
\includegraphics[scale=0.3]{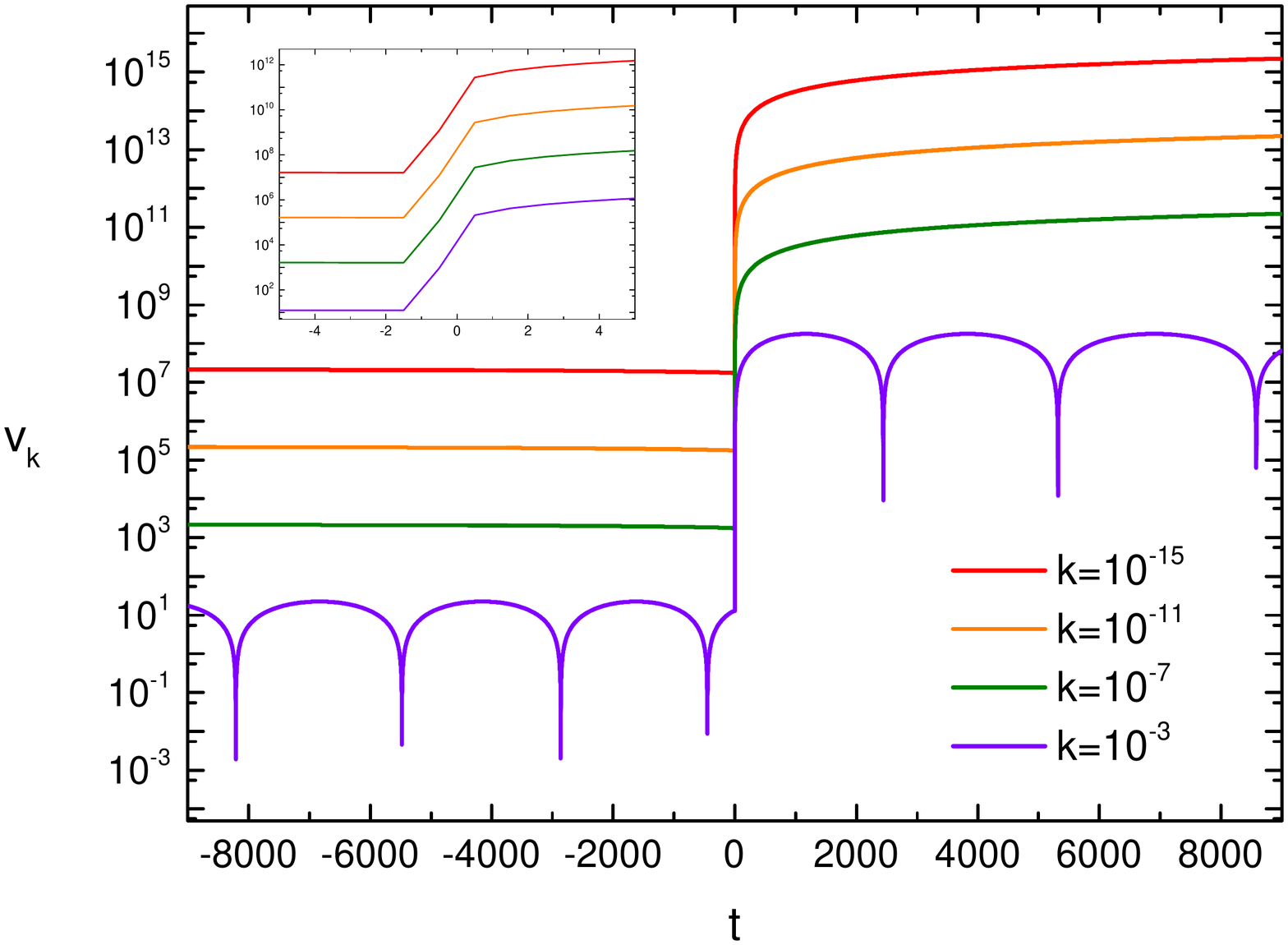}
\caption{Numerical plot of four groups of canonical perturbation modes $v_k$ (vertical axis)
as functions of cosmic time (horizontal axis). These modes are distinguished by the
comoving wave number $k$, which are $k=10^{-15}$ (in red), $k=10^{-11}$ (in orange),
$k=10^{-7}$ (in green), and $k=10^{-3}$ (in violet), respectively. The inner insert shows
the detailed evolution of $v_k$ during the bounce phase. The initial conditions of the
background field and background parameters are the same as those chosen in Fig. 2.
All numerical values are in Planck units $M_p$. The initial conditions for perturbation
modes were chosen as the vacuum initial condition given by Eq. (51). }
\label{Fig:vsk}
\end{figure}

First, we studied the time evolution of a set of canonical perturbation modes $v_k$
as a function of  cosmic time. The results are shown in Fig. \ref{Fig:vsk}. The
comoving wave numbers $k$ chosen are (in Planck units) $10^{-15}$, $10^{-11}$,
$10^{-7}$, and $10^{-3}$. First, we see from Fig. \ref{Fig:vsk} that the larger the
comoving wave number is, the later the perturbation mode exits the Hubble radius and
ceases oscillation (the purple mode is still oscillating).

When the modes pass through the bounce point, their  amplitude is amplified. For the
long wavelength modes the amplification factor is clearly independent of $k$, and the
amplitude of the amplification factor agrees well with our analytic estimate for ${\cal F}$
from Section III.B.2. From our graph, it is clear that the amplitude grows approximately
exponentially with cosmic time. This is due to the fact that the parameter $z$ undergoes
a rapid change. The equation of motion for gravitational waves is similar to that of
cosmological perturbations, except that the function $z(t)$ is to be replaced by the scale
factor $a(t)$. Since the scale factor is approximately constant around the bounce point,
the amplitude of gravitational waves does not jump at the bounce. Hence, the evolution
of fluctuations about the bounce point produces a natural suppression of the tensor
to scalar ratio of fluctuations by about a factor of ${\cal F}^2$ which obviates the need
for a matter bounce curvaton. This growth of scalar modes which we find is closely
related to the instability of nonsingular bounces discussed in the case of the
New Ekpyrotic scenario in \cite{BingXue}. We have shown that in our bouncing model
the instabilities of fluctuation modes remain under control and do not destroy the
predictions of the cosmological model.

\begin{figure}
\includegraphics[scale=0.3]{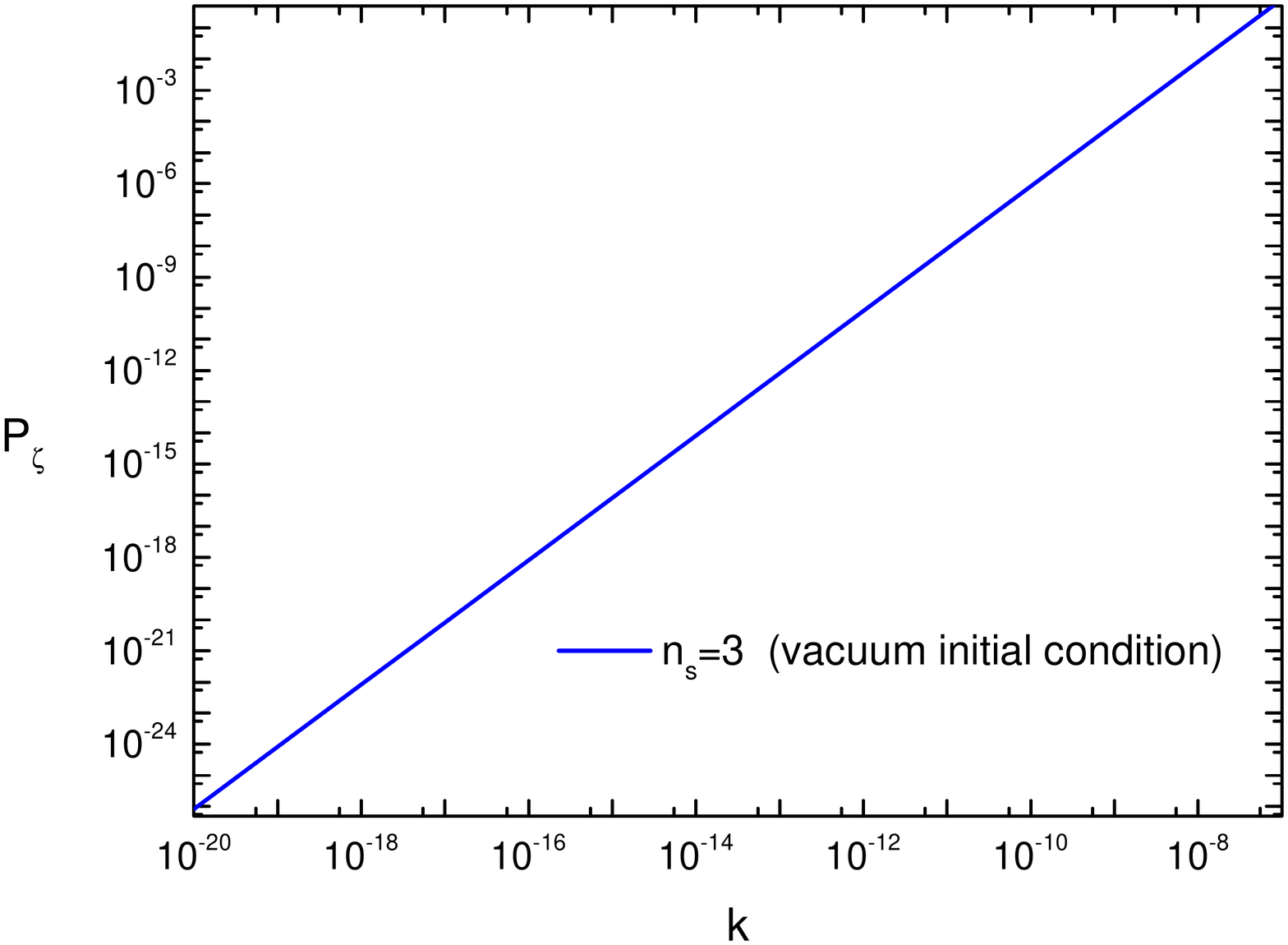}
\caption{Numerical plot of the primordial power spectrum $P_{\zeta}$ (vertical axis) as a
function of comoving wave number $k$ (horizontal axis). The initial conditions for the
background field and background parameters are the same as those chosen in Fig. 2, and
Planck units are used. The initial conditions for perturbation modes are the same as
those chosen in Fig. \ref{Fig:vsk}. The spectral index is indicated in the plot: $n_s\simeq3$. }
\label{Fig:Pzeta}
\end{figure}

In Fig. \ref{Fig:Pzeta}, we show the scale dependence of the primordial post-bounce
power spectrum  $P_{\zeta}$ as a function of the comoving wave number. The plot is
on a log-log scale to accommodate the wide range of scales. From the figure, we
can read off that $P_\zeta$ is a power law function of $k$ with the spectral index
$n_s\simeq3$ for perturbation modes directly generated from vacuum fluctuations.
This result is exactly the same as what we obtained in Eq. (56) and the discussion
performed there.

\section{Conclusions and Discussion}

In this paper we have presented a single scalar field model which yields a nonsingular bouncing cosmology. It makes use of a negative exponential potential which yields an Ekpyrotic contracting phase. At high energy densities, the scalar field undergoes a phase transition to a ghost condensate. This leads to the violation of the Null Energy Condition which is required to obtain a nonsingular bounce in flat FRW models in General Relativity. Following the bounce, a period of kinetic-driven expansion results.

By adding regular matter and radiation to the model, we obtain a realization of the ``matter bounce'' scenario which is free of the anisotropy problem which plagues other realizations of this scenario. The cosmological scenario is therefore as follows: the universe begins in a contracting phase with cold matter dominating over relativistic radiation and over the scalar field. Due to its equation of state $w_c > 1$ the scalar field comes to dominate the universe and leads to an Ekpyrotic-type phase of contraction. This phase of contraction is free from the BKL instability since the energy density in anisotropies grows less fast than that in the scalar field. This phase is followed by the ghost condensate-driven bounce which in turn ends in a kinetic-driven expanding period. Eventually, the usual radiation and matter come to dominate the energy density again, leading to the a Standard Big Bang expanding universe.

We have performed a detailed study of the evolution of cosmological fluctuations in our model. We have shown that the spectrum retains its slope through the bounce. Thus, vacuum fluctuations which exit the Hubble radius during the matter-dominated phase of contraction acquire and maintain a scale-invariant spectrum. Thus, our
model provides a realization of the ``matter bounce'' alternative to inflation which is free from the anisotropy problem which plagues previous realizations. We have also shown that in the absence of initial matter and radiation dominated phases of contraction, it is possible to obtain a scale-invariant spectrum of fluctuations from initial thermal particle inhomogeneities provided the equation of state during the contracting phase takes on a particular value.

We have found that the fluctuation modes undergo a period of exponential growth during the bounce phase. The growth factor ${\cal F}$ of the fluctuation mode during this phase, while being large in amplitude, is independent of $k$. Hence, the spectral shape passes through the bounce without change. The amplitude
of scalar modes increases relative to that of tensor modes (for which ${\cal F} \simeq 1$). Hence, a small tensor to scalar ratio results. The increase of the amplitude of the fluctuations during the bounce phase has another implication: the value of the Hubble constant at the transition between matter-dominated contraction and Ekpyrotic contraction can be fairly low. If we compare the final amplitude (\ref{amplitude}) of the curvature power spectrum with what is required to match observations, then with ${\cal F} \sim 1$ we would require a very large value of the Hubble expansion parameter at the beginning of the Ekpyrotic phase. In this context, the anisotropy problem might have reappeared: the initial anisotropies cannot be larger than a critical value such that they begin to dominate at the beginning of the Ekpyrotic phase.

In this paper we have focused on adiabatic fluctuations only. It would be interesting to study entropy modes in the model.

\begin{acknowledgments}

We thank Justin Khoury, Paul Steinhardt and BingKan Xue for stimulating discussions. The work of YFC and DAE is supported in part by the DOE and by the Cosmology Initiative at Arizona State University. RB is supported in part by an NSERC Discovery Grant and by funds from the Canada Research Chair program. He was also supported by a Killam Research Fellowship.

\end{acknowledgments}

\section{Appendices}

In the first part of this Appendix, we study the general treatment of linear cosmological perturbations within the KBG model. In the second part, we show how a Fourier mode of cosmological perturbations is transferred from the contracting phase to the bouncing phase and then to the expanding period through the matching condition.

\subsection{The derivation of the quadratic action of cosmological perturbations}

It is useful to study perturbation theory by making use of the ADM metric:
\begin{eqnarray}\label{ADM_metric}
 ds^2 = N^2dt^2 - h_{ij}(dx^i+N^idt)(dx^j+N^jdt)~,
\end{eqnarray}
where $N$ and $N^i$ are the lapse function and shift vector, respectively. Making use of this metric (\ref{ADM_metric}), one can decompose the original action of the model minimally coupled to Einstein gravity into time and space parts. To be explicit, the action can be written as
\begin{eqnarray}\label{ADM_action}
 S &=& \int dt d^3x \sqrt{h} \frac{N}{2} \bigg[ M_p^2(R_3+\kappa_{ij}\kappa^{ij}-\kappa^2) \nonumber\\
 && + K(\phi, X) + G(\phi, X)\Box\phi \bigg]~,
\end{eqnarray}
where $R_3$ is the Ricci scalar defined on the three-dimensional space and
\begin{eqnarray}
 \kappa_{ij} \equiv \frac{1}{2N}(\dot{h}_{ij}-\nabla_iN_j-\nabla_jN_i)
\end{eqnarray}
is the extrinsic curvature. Since we want to investigate the cosmological perturbations which vary in space and time, one has to include the space-dependence in the expression for the kinetic term of the scalar $\phi$ which is given by
\begin{eqnarray}
 X = \frac{1}{2N^2}(\dot\phi-N_ih^{ij}\partial_j\phi)^2 - \frac{1}{2}h^{ij}\partial_i\phi\partial_j\phi~.
\end{eqnarray}

We are interested only in the part of the action involving scalar metric and matter fluctuations. It is well known that, for a single scalar matter field minimally coupled to Einstein gravity, there exists only one scalar type of degree of freedom. We choose the uniform field gauge
\begin{eqnarray}\label{gauge pert}
 \delta\phi = 0~,~~ h_{ij} = a^2 e^{2\zeta}\delta_{ij}~,
\end{eqnarray}
so that the linear perturbations of the scalar field in the field Lagrangian can be eliminated. Further, the scalar components of the lapse function and shift vector can be determined through the Hamiltonian and Momentum constraints. Explicitly, the scalar contributions to the lapse function and shift vector take the form
\begin{eqnarray}\label{alphasigma}
 N = 1+\alpha~,~~N_i = \frac{\partial_i\sigma}{M_p}~,
\end{eqnarray}
up to leading order. One can insert Eqs. (\ref{gauge pert}) and (\ref{alphasigma}) into the action (\ref{ADM_action}) and then expand it to quadratic order. After a lengthy calculation, the form of the second order action is written as,
\begin{eqnarray}\label{ADM_S2}
 S_2 &=& \int dt d^3x a^3 \bigg[ (2M_p^2\dot\zeta-2M_p^2H\alpha+\dot\phi^3G_{,X}\alpha)\frac{\partial_i^2\sigma}{M_pa^2} \nonumber\\
 &&-3M_p^2\dot\zeta^2 - 2M_p^2\alpha\frac{\partial_i^2\zeta}{a^2} + 6M_p^2H\alpha\dot\zeta - 3\dot\phi^3G_{,X}\alpha\dot\zeta \nonumber\\
 && + M_p^2\frac{(\partial_i\zeta)^2}{a^2} - 3M_p^2H^2\alpha^2 + \frac{\dot\phi^2}{2}K_{,X}\alpha^2 \nonumber\\
 && + \frac{\dot\phi^4}{2}K_{,XX}\alpha^2 + 6H\dot\phi^3G_{,X}\alpha^2 + \frac{3}{2}H\dot\phi^5G_{,XX}\alpha^2 \nonumber\\
 &&- \dot\phi^2(G_{,\phi}+\frac{\dot\phi^2}{2}G_{,X\phi})\alpha^2
 \bigg]~,
\end{eqnarray}
which involves the perturbation variables $\zeta$, $\alpha$ and $\sigma$. Varying the action (\ref{ADM_S2}) with respect to $\alpha$ and $\sigma$ yields the following relations,
\begin{eqnarray}
\label{alpha}
 \alpha &=& \frac{2M_p^2\dot\zeta}{2M_p^2H-\dot\phi^3G_{,X}}~, \\
\label{sigma}
 \partial_i^2\sigma &=& 3M_pa^2\dot\zeta - \frac{2M_p^3\partial_i^2\zeta}{2M_p^2H-\dot\phi^3G_{,X}} \nonumber\\
 && + \frac{2M_p^3a^2\dot\zeta(-6M_p^2H^2 +\dot\phi^2K_{,X} +\dot\phi^4K_{,XX})}{(2M_p^2H-\dot\phi^3G_{,X})^2} \nonumber\\
 && + \frac{2M_p^3a^2\dot\zeta(12H\dot\phi^3G_{,X} +3H\dot\phi^4G_{,XX})}{(2M_p^2H-\dot\phi^3G_{,X})^2} \nonumber\\
 && - \frac{2M_p^3a^2\dot\zeta(2\dot\phi^2G_{,\phi} +\dot\phi^4G_{,X\phi})}{(2M_p^2H-\dot\phi^3G_{,X})^2}~,
\end{eqnarray}
which are exactly the Hamiltonian and Momentum constraints.

Finally, by making use of the relations (\ref{alpha}), (\ref{sigma}) and the quadratic action (\ref{ADM_S2}), we can obtain the action of scalar perturbation at linear order, whose form is
\begin{eqnarray}\label{S2_t}
 S_2 = \int dt d^3x  \frac{a(t)}{2} z^2(t) \bigg[\dot\zeta^2 - \frac{c_s^2(t)}{a^2(t)}(\partial_i\zeta)^2 \bigg]~,
\end{eqnarray}
where we have introduced a function $z$ which is completely determined by the background evolution. Its explicit form is given by
\begin{eqnarray}\label{z2_general}
 z^2 = \frac{2M_p^4a^2\dot\phi^2{\cal P}}{(2M_p^2H-\dot\phi^3G_{,X})^2}~,
\end{eqnarray}
with ${\cal P}$ being mentioned in Eq. (\ref{Pterm}), which is the coefficient of the second order derivative term of the background equation of motion. From Eq. (\ref{z2_general}), one can immediately find that the positivity of the coefficient ${\cal P}$ directly determines the positivity of $z^2$ and thus can be used to judge whether there is a ghost mode or not. The square of the sound speed is given by
\begin{eqnarray}\label{cs2_general}
 c_s^2 &=& \frac{1}{{\cal P}} [ K_{,X} +4H\dot\phi{G}_{,X} -\frac{\dot\phi^4G_{,X}^2}{2M_p^2} -2G_{,\phi} \nonumber\\
 && +\dot\phi^2G_{,X\phi} +(2G_{,X}+\dot\phi^2G_{,XX})\ddot\phi ]~.
\end{eqnarray}

\subsection{Details of the matching condition calculation}

We first match the cosmological perturbation $v_k^c$ and $v_k^b$ at the moment $\tau_{B-}$, and then determine the coefficients $d_1$ and $d_2$ as follows
\begin{eqnarray}
 &d_1 \simeq -\frac{ c_2 \Gamma_{\nu_c} e^{\omega(\tau_B-\tau_{B-})} [1-2\nu_c+2\omega(\tau_{B-}-\tilde\tau_{B-})] }{ 2^{2-\nu_c} \pi \omega k^{\nu_c} (\tau_{B-}-\tilde\tau_{B-})^{\nu_c+\frac{1}{2}} } \\
 &d_2 \simeq \frac{ c_2 \Gamma_{\nu_c} e^{-\omega(\tau_B-\tau_{B-})} [1-2\nu_c-2\omega(\tau_{B-}-\tilde\tau_{B-})] }{ 2^{2-\nu_c} \pi \omega k^{\nu_c} (\tau_{B-}-\tilde\tau_{B-})^{\nu_c+\frac{1}{2}} }~,
\end{eqnarray}
in which the dominant contribution comes from the $c_2$ mode of the perturbation in the contracting phase.

Similarly, we match the perturbations $v_k^b$ and $v_k^e$ at the end of the bouncing phase $\tau_{B+}$ and determine the coefficients $e_1$ and $e_2$. We make use of these coefficients and then can extract the dominant mode of $v_k^e$ as follows,
\begin{eqnarray} \label{form}
 v_k^e(\tau) &\simeq& \frac{ {\cal F} (\gamma_E+\ln\frac{{\cal H}_{B+}}{\cal H}) \Gamma_{\nu_c} 
  }{ 2^{2-\nu_c} \pi (\tau_{B-}-\tilde\tau_{B-})^{\nu_c-\frac{1}{2}} } \nonumber\\
 && \times c_2(k) k^{-\nu_c} \bigg(\frac{\tau-\tilde\tau_{B+}}{\tau_{B+}-\tilde\tau_{B+}}\bigg)^{\frac{1}{2}}~.
\end{eqnarray}
Note that the boost factor ${\cal F}$ and the term inside the square brackets appearing in the numerator of this formula give the amplification factor which arises from the tachyonic instability when the perturbation mode evolves through the bouncing phase. This feature is quite interesting for phenomenological studies of cosmological perturbations in bouncing cosmologies. For example, in \cite{bouncecurvaton} a bounce curvaton
scenario was proposed based on a tachyonic amplification of primordial isocurvature perturbations which is very similar to the tachyonic instability encountered here. The same process can help to improve the efficiency of preheating in a bouncing universe \cite{Cai:2011ci}. Because of the large value of the amplification factor which we find in our current model, we do not need to invoke a bounce curvaton mechanism to suppress
the tensor modes relative to the scalar one.

\end{document}